\documentclass[twocolumn,10pt]{asme2e}

\usepackage{epsfig} 
\usepackage{graphicx}
\usepackage{float}
\usepackage{listings}
\usepackage{multirow}
\usepackage{mathtools}
\usepackage{breqn}
\usepackage{cite}
\usepackage{cleveref}
\usepackage{amsmath} 
\usepackage{amssymb}  
\usepackage{amsthm}
\usepackage{subfig, epstopdf, graphicx}
\usepackage{algorithm, algorithmicx}
\usepackage{algpseudocode}
\usepackage{color}

\graphicspath{{./pdffigure/}}
\confshortname{DSCC 2019}
\conffullname{ASME 2019 Dynamic Systems and Control Conference}
\confdate{October 8-11}
\confyear{2019}
\confcity{Park City, Utah}
\confcountry{USA}

\papernum{DSCC2019-9015}

\title{Magnetically Actuated Artificial Microswimmers As Mobile Microparticle Manipulators}

\author{Jake Buzhardt
    \affiliation{
	Department of Mechanical Engineering\\
	Clemson University\\
	Clemson, S.C, 29632\\
    Email: jbuzhar@g.clemson.edu
    }
}

\author{Phanindra Tallapragada 
    \affiliation{
	Department of Mechanical Engineering\\
	Clemson University\\
	Clemson, S.C, 29632\\
    Email: ptallap@clemson.edu
    }
}

\begin{document}

\maketitle

\begin{abstract}
{\it Micro-scale swimming robots have been envisaged for many medical applications such as targeted drug delivery, where the microrobot will be expected to navigate in a fluid through channels carrying a payload. 
Alternatively, in many cases, such a payload does not have to be physically bound to the swimmer, but may be  instead manipulated and steered through the channel by the microrobot. 
We investigate this problem of contactless manipulation of a microparticle by mobile microswimmer in a fluid at low Reynolds number. 
We consider a model of a magnetically actuated artificial microswimmer, whose locomotion through a fluid induces a disturbance velocity field in the fluid, that then acts to propel a cargo particle in its vicinity. 
The problem investigated in this paper is therefore one of coupled locomotion-manipulation of two bodies in a fluid. The magnetic swimmer's motion is actuated by an externally applied magnetic field of constant strength but whose direction rotates at a constant rate in a plane. 
The swimmer propels itself in the direction perpendicular to this plane if the frequency associated with the periodic magnetic field is above a critical frequency. Below this critical frequency, the swimmer tumbles in place without net locomotion. 
The coupled fluid-swimmer-cargo particle dynamics are solved numerically using the method of Stokesian dynamics. 
The induced motion of the cargo particle is shown to be controllable. 
This is achieved by switching the planes of rotation of the magnetic field and switching frequency of the magnetic field above and below the critical frequency. 
While a swimmer with a specific geometry has been used in the model, the results of this paper are applicable to swimmers with other geometries and means of propulsion. 
The results of this paper show that microswimmers can be utilized as mobile manipulators of microparticles in a fluid.
}
\end{abstract}

\section{Introduction}
Microrobots that can swim and navigate in microchannels hold great potential for many medical applications such as targeted drug delivery, cellular manipulation and diagnostics \cite{nelson_arbe_2010, nelson_cts_2017, asen_nano_2013, 2018RAL_Hunter}. 
Such microrobots can derive their propulsive ability via magnetic\cite{nelson_apl_2009, kim_pre_2014, nelson_cts_2017}, electric\cite{kuhn_natcomm_2011}, acoustic \cite{crespi_srep_2015}, optical\cite{palffy_nature_2004} or chemical\cite{ozin_nano_2010, solovev_nano_2012} actuation. 
These microrobots are expected to carry a payload or cargo that by itself cannot be propelled or steered very well and require that the cargo particles be bonded to the microrobot and when required released by the robot. 
This can pose limitations on the cargo that can be carried. 
An alternative approach that can overcome these limitations is via the contactless manipulation of cargo particles by a microrobot. In essence the microrobot becomes a mobile contactless micromanipulator. 

\begin{figure}[ht]
    \centering
    \includegraphics[width =0.7\linewidth]{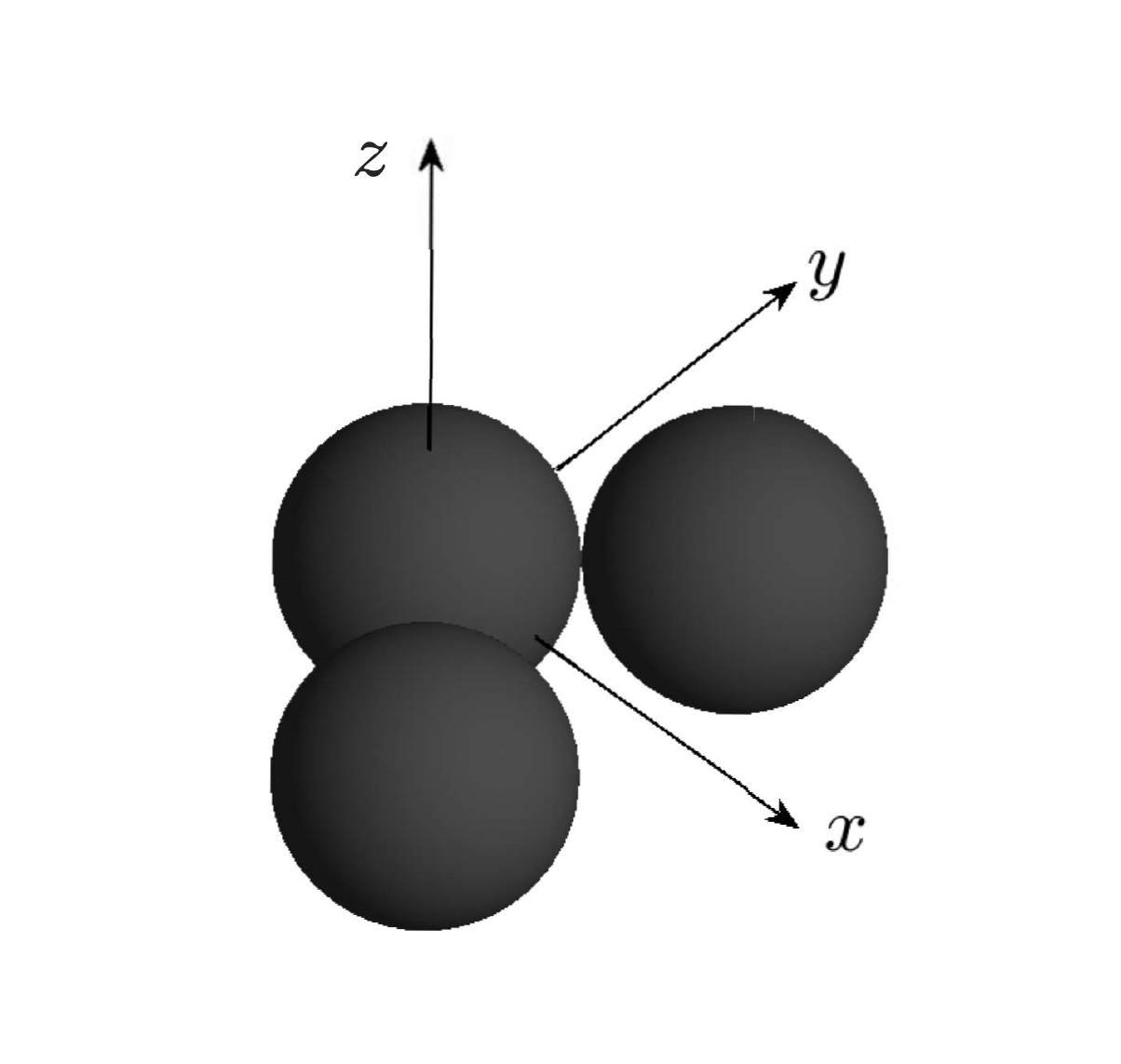}
    \caption{Three-sphere artificial microswimmer considered in this work, shown in the axes of the body-fixed frame of reference.}
    \label{fig:Swimmer}
\end{figure}

In this paper we investigate through computations the ability of a magnetically actuated microswimmer to steer a non-magnetic particle in its neighborhood in a contactless manner. 
The microswimmer geometry considered here consists of three rigid magnetic spheres bonded together, with the centers of the spheres forming the vertices of a triangle, as shown in Fig. \ref{fig:Swimmer}. 
Such a geometry has been proposed to be the simplest body capable of propulsion in a low Reynolds number fluid environment, the propulsion being enabled by the achiral geometry of the swimmer \cite{kim_pre_2014}. 
Magnetic swimmers such as this one can be propelled using a magnetic field of constant strength whose direction rotates periodically in a plane. 
When the frequency of the rotating magnetic field is below a value called the step-out frequency, $\omega_s$, the angle between the magnetic moment of the swimmer and the periodically rotating magnetic field vector converges to a constant value. 
This induces a constant torque on the swimmer. 
The achirality of the swimmer leads to a coupling of this torque to the velocity of the body, propelling the swimmer in a direction perpendicular to the plane of the magnetic field. 
The swimmer can be steered in different directions by changing the plane of rotation of the magnetic field \cite{kim_feedback2015}.  

The fluid environment in which the microswimmer moves is dominated by viscous dissipative effects while the inertia of both the swimmer and the fluid are negligible. 
Thus, the fluid dynamics of such low Reynolds number fluid flows are well-described by the Stokes equations. 
In this setting, the forces acting on the body are directly proportional to the velocity of the body. 
This linear relationship between the velocity of a single swimmer or body and the forces acting on it is modeled by the so called mobility matrix \cite{happel_brenner, kim_karrila}.
The dynamics of a single swimmer in the absence of any other  particles or other boundaries are therefore entirely determined by its mobility matrix if the forces and moments acting on the body are known. 
However, when the swimmer is close to other boundaries, such as other particles, the dynamics of the swimmer-particle interaction must be modeled via the so called grand mobility tensor \cite{happel_brenner, kim_karrila}. 
The grand mobility tensor couples the kinematics of the two bodies and the forces acting on them.
To compute this grand mobility tensor we adopt the well known Stokesian Dynamics method \cite{brady_afm_1988}, which is very well suited to the study the dynamics of a swimmer that can be discretized as spheres. 

In this paper, we theoretically model the manipulation of a spherical cargo particle by a magnetized three-sphere swimmer where the length scales of the cargo particle and the swimmer are comparable. 
The equations governing this interaction form a driftless control affine system.
The driftless nature of the control system is a consequence of the lack of inertia in a Stokes flow. 

We show that the swimmer may be used as a mobile microparticle manipulator steering the non-magnetic cargo particle in a desired direction, as steering the swimmer effectively controls the fluid velocity field experienced by the cargo particle. This observation is then utilized to determine a magnetic field input that can generate motion of both the magnetic swimmer and the passive particle in arbitrarily chosen directions. 

\section{Modeling of the swimmer-particle interaction}

\subsection{Hydrodynamic Modeling}
The motion of the artificial microswimmers and particles considered in this work are described by very small length and velocity scales.  
In such a setting, the fluid dynamics generated by the particles is dominated by viscous dissipation and inertial effects are considered negligible.
This fluid motion is governed by the Stokes equations, given by  
\begin{subequations}
\begin{align}
    -\nabla p + \mu \nabla^2\mathbf{u} = \mathbf{0}\\
    \nabla \cdot \mathbf{u} = 0
    \end{align}
\end{subequations}
where $\mathbf{u}$ and $p$ are the fluid velocity and pressure fields respectively, and $\mu$ is the fluid viscosity.  

The linearity of the Stokes equations implies that the forces and torques operating on a rigid body moving through a viscous fluid are linearly related to the velocity and angular velocity of the body \cite{happel_brenner}.
This relationship is known as the mobility relationship and is given here by 
\begin{equation}
    \begin{pmatrix}
    \mathbf{V}\\
    \boldsymbol{\Omega}
    \end{pmatrix}
    =
    \begin{pmatrix}
    \mathbf{K}  &   \mathbf{C}\\
    \mathbf{C}^{\intercal}    &   \mathbf{M}
    \end{pmatrix}
    \begin{pmatrix}
    \mathbf{F}^e\\
    \mathbf{T}^e
    \end{pmatrix}
\end{equation}
where $\mathbf{F}^e$ and $\mathbf{T}^e$ are the external forces and torques acting on the body and $\mathbf{V}$ and $\boldsymbol{\Omega}$ are the velocity and angular velocity of the body.
The self-mobilities,  $\mathbf{K}$, $\mathbf{C}$, and $\mathbf{M}$ are $3\times3$ geometry dependent matrices that describe the hydrodynamic effects on the body's motion.  The mobility matrix associated with this particular geometry has been given previously by Ref. \cite{fu_pre_2014,Leshansky_prf_2017}.

The geometry of the artificial swimmer considered in this paper consists of three spheres, rigidly connected with a $90^{\circ}$ bend.  
This body is chosen both for the structure of its mobility matrix and for its ease of fabrication.  
The swimmer is to be driven by a torque produced by a magnetic field, with no external forcing $\mathbf{F}=\mathbf{0}$. 
Thus, to achieve a translational motion, it is necessary that the matrix coupling torque to translational velocity be nonzero, $\mathbf{C} \neq \mathbf{0}$.  
One of the simplest geometric conditions for a body to have a nonzero $\mathbf{C}$ matrix is that the body have two mutually perpendicular planes of symmetry, but not three \cite{happel_brenner}. 

In cases where multiple bodies are present in a fluid domain, it is also necessary to consider the hydrodynamic coupling between bodies. 
The relationship between the forcings and the body kinematics is still linear in this case, and is represented using a grand mobility matrix. 
In this work, we consider the motion of particle in the presence of an externally driven artificial microswimmer, so the grand mobility relationship is given by 

\begin{equation}
    \begin{pmatrix}
    \mathbf{V}_s\\[0.5ex]
    \mathbf{V}_p\\[0.5ex]
    \boldsymbol{\Omega}_s\\[0.5ex]
    \boldsymbol{\Omega}_p
    \end{pmatrix}
    =
    \begin{pmatrix}
    \mathbf{K}_{ss}  &   \mathbf{K}_{sp}    &   \mathbf{C}_{ss}    &   \mathbf{C}_{sp}\\[0.5ex]
    \mathbf{K}_{ps}  &   \mathbf{K}_{pp}    &   \mathbf{C}_{ps}    &   \mathbf{C}_{pp}\\[0.5ex]
    \mathbf{C}_{ss}^{\intercal} &   \mathbf{C}_{sp}^{\intercal}    &   \mathbf{M}_{ss}  &   \mathbf{M}_{sp}\\[0.5ex]
    \mathbf{C}_{ps}^{\intercal} &   \mathbf{C}_{pp}^{\intercal}    &   \mathbf{M}_{ps}  &   \mathbf{M}_{pp}\\[0.5ex]
    \end{pmatrix}
    \begin{pmatrix}
    \mathbf{F}_s^e\\[0.5ex]
    \mathbf{F}_p^e\\[0.5ex]
    \mathbf{T}_s^e\\[0.5ex]
    \mathbf{T}_p^e
    \end{pmatrix}
    \label{eq:GrandMob}
\end{equation}
where the subscripts $s$ and $p$ indicate the particle and the swimmer respectively. 
In the case of a single swimmer, it is oftentimes convenient to represent these dynamics in a body-fixed frame of reference, as the sub-matrices $\mathbf{K}$, $\mathbf{C}$, and $\mathbf{M}$ are dependent only on the body geometry and thus remain constant in a body-fixed frame.  
However, in the case of multiple interacting bodies, each of the terms in this matrix depend on the position and orientation of each body relative to one another.  
Therefore, it is more convenient to define these matrices in a common, spatially-fixed reference frame. 

To define the grand mobility matrix for this system, we implement the commonly-used Stokesian dynamics algorithm.  
The method is especially suited for systems composed of spheres moving in a viscous fluid.  
This method relies on the fact that the fluid velocity field generated can be written (in tensor notation) by expanding the velocity field in moments about the center of each particle as 
\begin{equation}
\begin{split}
    u_i(x) = \frac{1}{8\pi\mu}\Bigg[ \bigg(1+\frac{a^2}{6}&\nabla^2 \bigg)J_{ij} \, F_j + R_{ij}\, T_j \\
    &+ \left(1+\frac{a^2}{10}\nabla^2 \right)K_{ijk}\, S_{jk} + \cdots \Bigg]
    \end{split}
\end{equation}
where $J_{ij}$, $R_{ij}$ and $K_{ijk}$ are the propagators associated with the singularities of Stokes flow, the Stokeslet, rotlet, and stresslet respectively \cite{brady_afm_1988}. 
Also, in this equation, $F_j$, $T_j$, and $S_{jk}$ represent the force, torque, and stress components resulting from the motion of the sphere, and $a$ is the particle radius.  
This result, along with the well-known Fax\'en relationships are used to determine the motion of a collection of spheres.  
That is, this may be used to generate a grand mobility matrix that describes the hydrodynamic coupling between all spheres in a system.
Full details of this derivation along with the expressions for the elements of this many-sphere grand mobility matrix are given in Ref \cite{durlofsky_brady_bossis_1987}.

To condense this matrix describing the motion of all spheres (denoted by $\alpha$) into a matrix describing the motion of a rigid body composed of spheres (denoted by $A$), we apply the conditions of rigid body motion that 
\begin{subequations}
\begin{align}
    \mathbf{V}_{\alpha} &= \mathbf{V}_A + \boldsymbol{\Omega}_A\times \mathbf{r}_{\alpha A}\\
    \boldsymbol{\Omega}_{\alpha} &= \boldsymbol{\Omega}_A 
    \end{align}
\end{subequations}
and the quasi-static conditions of rigid body motion in a Stokes flow 
\begin{subequations}
\begin{align}
    \mathbf{F}^e-\sum_{\alpha}\mathbf{F}_{\alpha} &= \mathbf{0}\\
    \mathbf{T}^e-\sum_{\alpha}\mathbf{T}_{\alpha} &= \mathbf{0}.
\end{align}
\end{subequations}
Here $\mathbf{r}_{\alpha A}$ is the vector from a chosen center of the rigid body $A$ to the center of the sphere $\alpha$. 
In all simulations herein, the center of rigid body is chosen to correspond with the center of hydrodynamic mobility, as defined in \cite{Leshansky_prf_2017}.
Full details of this process are found in \cite{SwanBrady_Teaching}.  
The result of the process outlined here is a grand mobility matrix of the same form as shown in Eq. \eqref{eq:GrandMob} that gives the hydrodynamic coupling between the artificial microswimmer and the spherical particle.  
Since this matrix is dependent on the position and orientation of both bodies, in simulation, it is updated at each time-step. 

\subsection{Rigid Body Dynamics}
In this work, we consider the propulsion of a body by an externally applied magnetic field.  
For this, the magnetization of the body is described by a magnetic moment vector, which is assumed to remain constant in a body-fixed frame of reference.  
The externally applied torque acting on the body of the artificial swimmer is then given by 
\begin{equation}
    \mathbf{T}_s^e = (\mathbf{R}^{\intercal}\,\mathbf{m})\times\mathbf{B}
    \label{eq:MagTorque}
\end{equation}
where $\mathbf{B}$ is a vector representing the externally applied magnetic field, defined in a spatially-fixed frame of reference, $\mathbf{m}$ is the magnetic moment vector, defined in a body-fixed frame of reference, and $\mathbf{R}^{\intercal}$ is the matrix giving the rotation transformation from the body-fixed frame to the spatially fixed frame of reference. Thus, the torques and velocities are also considered in the spatially-fixed frame of reference. 
The torque $\mathbf{T}^e_s$ is the only external actuation imparted to the system, as no forces are applied and the spherical particle is not magnetized, so it is unaffected by the magnetic field. 

Since the dynamics are dependent on the body orientation relative to the spatially-fixed reference frame, it is necessary to parameterize this orientation of the body.  
For this, we adopt the standard $ZXZ$ Euler angles.  
Thus, the transformation $\mathbf{R}$ from the spatially fixed frame to the body-fixed frame is given by 
\begin{equation}
    \mathbf{R} = \mathbf{R}_z(\psi)\,\mathbf{R}_x(\theta)\,\mathbf{R}_z(\phi) 
    \label{eq:Rotation}
\end{equation}
where $\mathbf{R}_x$ and $\mathbf{R}_z$ are the matrices representing the rotation by the given angles about the body $x$ and $z$ axes respectively.  
Since these matrices are orthogonal, the inverse transformation from the body-fixed frame to the spatially-fixed frame is given by $\mathbf{R}^{-1}=\mathbf{R}^{\intercal}$, as in Eq. \eqref{eq:MagTorque}, where the superscript $\intercal$ indicates matrix transposition. 

With this parameterization, the time evolution of the Euler angles can be related to the angular velocities in a body-fixed frame of reference as 
\begin{equation}
    \boldsymbol{\Omega}_b = 
    \begin{pmatrix} 
    \sin{\psi}\sin{\theta}  &   \cos{\psi}  &   0\\[0.5ex]
    \cos{\psi}\sin{\theta}  &   -\sin{\psi}  &   0\\[0.5ex]
    \cos{\theta}    &   0   &   1
    \end{pmatrix}
    \begin{pmatrix}
    \dot{\phi}\\[0.5ex]
    \dot{\theta}\\[0.5ex]
    \dot{\psi}
    \end{pmatrix}
    \label{eq:AngleRates}
\end{equation}
where the dot notation indicates the time rate of change \cite{goldstein}.

\subsection{Control System Formulation}
By applying Eqs. (\ref{eq:MagTorque}-\ref{eq:AngleRates}) in conjunction with the grand mobility matrix derived from the Stokesian dynamics method and shown in Eq. \eqref{eq:GrandMob}, we have defined a nonlinear dynamical control system of the form 
\begin{equation}
    \frac{d\boldsymbol{\xi}}{dt} = \mathbf{f}(\boldsymbol{\xi},\boldsymbol{\upsilon})
\end{equation}
where the state vector $\boldsymbol{\xi}$ is 
\begin{equation}
  \boldsymbol{\xi} \triangleq 
  \begin{pmatrix} \mathbf{X}_s   \\[0.5ex]
  \mathbf{X}_p    \\[0.5ex]
  \boldsymbol{\Theta}_s \\[0.5ex]
  \boldsymbol{\Theta}_p \end{pmatrix} 
\end{equation}
where the vectors $\mathbf{X}$ represent the respective positions of the artificial swimmer and the passive particle and the vectors $\boldsymbol{\Theta}$ are vector representations of the $ZXZ$ Euler angles, all defined relative to a spatially-fixed frame of reference. 

For this system, the control input $\boldsymbol{\upsilon}$ can be taken to be the components of the magnetic field $\boldsymbol{\upsilon} = \mathbf{B}$.  However, it has previously been shown both experimentally and theoretically that propulsion of the artificial swimmer can be readily achieved by applying a magnetic field of the form
\begin{equation}
    \mathbf{B}_0(t) =
    B \cdot
    \begin{pmatrix} 
    \cos{\omega\,t}, &   \sin{\omega\,t}, &   0
    \end{pmatrix}
    ^{\intercal}
    \label{eq:MagField}
\end{equation}
 defined in the spatially-fixed frame of reference.  
 This equation represents a magnetic field rotating steadily about the spatial $z$-axis, which causes the swimmer to translate along this axis in either the positive or negative $z$ directions, depending on the frequency of magnetic field rotation, $\omega$. 
 However, in this work, we seek to show that the swimmer (and the passive particle) can be steered along any arbitrary direction.  
 So, we modify the form of Eq. \eqref{eq:MagField} and take the applied magnetic field to be of the form 
 \begin{equation}
    \mathbf{B}(t) =
    \mathbf{R}_l(\gamma)\cdot
    B \cdot
    \begin{pmatrix} 
    \cos{\omega\,t}, &   \sin{\omega\,t}, &   0
    \end{pmatrix}
    ^{\intercal}
    \label{eq:MagFieldRotY}
 \end{equation}
where the matrix $\mathbf{R}_l(\gamma)$ is a rotation about an axis $l$ by an angle $\gamma$. With out loss of generality, in this paper, we consider only the case where the axis $l$ is chosen to be the spatially-fixed $y$-axis. This allows us to direct the magnetic field along any vector in the $xz$-plane. 
With this, we  take the control inputs to be 
\begin{equation}
    \boldsymbol{\upsilon} \triangleq
    \begin{pmatrix}
    \gamma , & \omega
    \end{pmatrix}^{\intercal},
\end{equation}
the angle of rotation, $\gamma$ of the oscillating field about the $y$-axis and the rotation frequency, $\omega$ of the magnetic field. 



\section{Simulation Results}
With the particle dynamics well defined and the control system formulated, we are able to simulate the dynamics of the swimmer-particle system.  Starting from an initial state the system, the dynamics are integrated forward in time using \texttt{MATLAB}'s built in variable order differential equation solver, \texttt{ode113}. At each integration time-step, the dynamics are computed by the following algorithm:
\begin{enumerate}
    \item Positions of all spheres in the simulation are computed based on the body positions, orientations, and compositions.
    \item The Stokesian dynamics algorithm is applied to form a many-sphere grand mobility matrix. 
    \item The quasi-static conditions and constraints of rigid body motion are applied to condense this matrix to a many-body grand mobility matrix.
    \item The magnetically induced torque acting on the swimmer is computed using \eqref{eq:MagTorque}, by rotating the constant magnetic moment vector from the body-fixed frame to the spatially-fixed frame of reference. 
    \item The grand mobility matrix is used, along with the externally appilied forces and torques to compute the instantaneous velocity and angular velocity of each body. 
    \item The angular velocity of each body is related to the time rate of change of the corresponding $ZXZ$ Euler angles using  \eqref{eq:AngleRates}.
\end{enumerate}

The numerical values of the parameters used for the simulations shown in this work are outlined as follows.  The magnetic moment vector is defined in the body-fixed frame of reference as 
\begin{equation}
    \mathbf{m} = m\begin{pmatrix} 0\quad &   \frac{\sqrt{2}}{2}\quad& \frac{\sqrt{2}}{2} \end{pmatrix}^{\intercal}
\end{equation}
with a magnitude $m = 4.0 \times 10^{-15}$ N\,m/T. The strength of the magnetic field is held constant at $B = 5.0\times10^{-3}$ T. 
The artificial microswimmer is composed of rigidly linked magnetic spheres of radius $a = 2.25\, \mu$m, with an internal bend angle of $90^{\circ}$.  
Passive spheres considered in these simulations are also of radius $a = 2.25\, \mu$m. 
These numerical values are chosen based on values reported in the experimental works on this subject, specifically \cite{kim_pre_2014}.

\begin{figure}
    \centering
    \includegraphics[width=\linewidth]{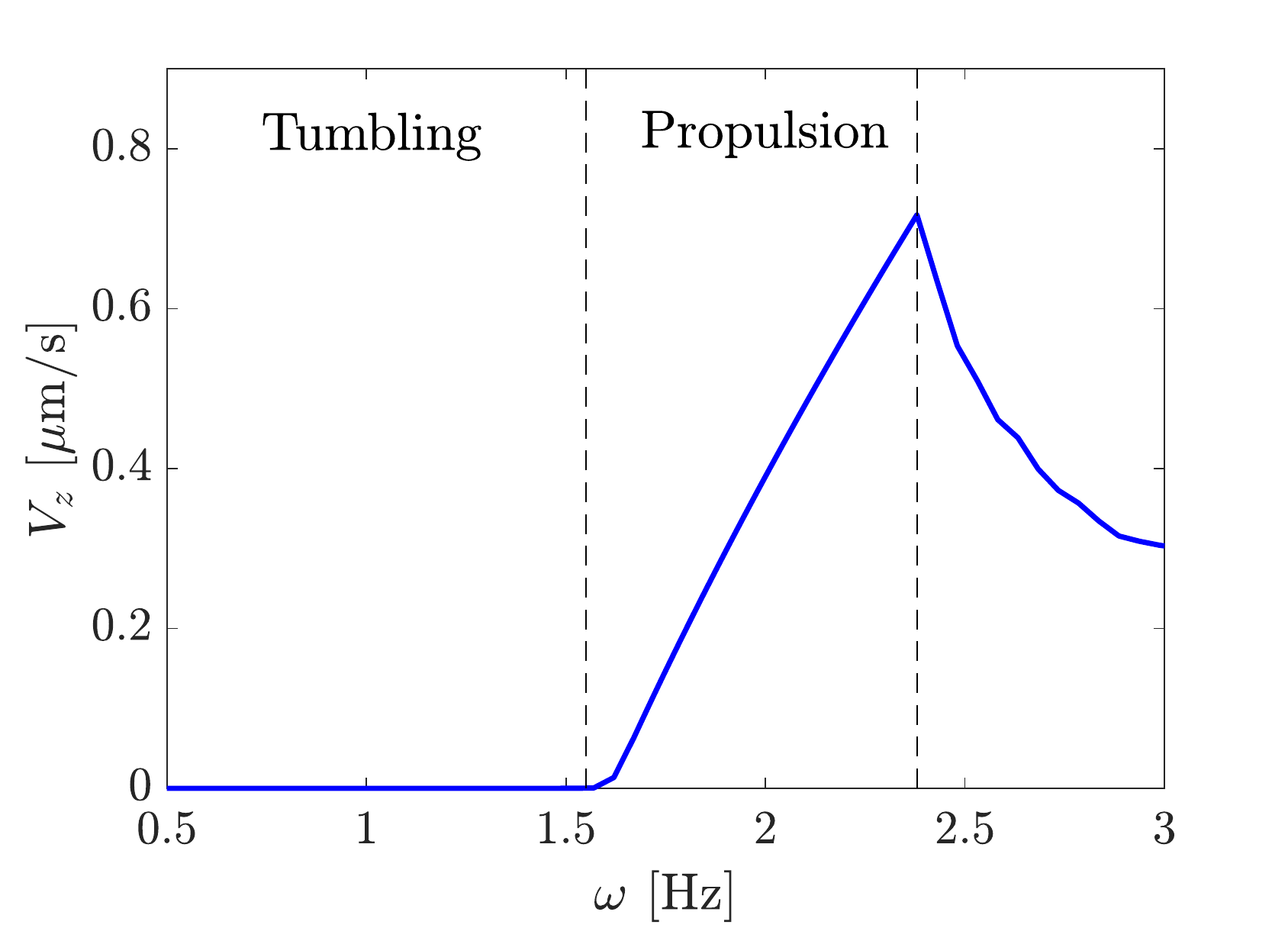}
    \vspace{1ex}
    \caption{Dependence of swimmer propulsion velocity on driving frequency $\omega$ of the rotating magnetic field. }
    \label{fig:VelFreq}
\end{figure}

Before considering the dynamics of swimmer-particle interaction, we briefly explain the dynamics of a single swimmer, propelled by a rotating magnetic field.  
It is a well known result that magnetically actuated artificial swimmers only exhibit meaningful locomotion for a certain range of driving frequencies of the oscillating field \cite{GhoshPRE,Nelson_Nano2010,ManLauga}. 
Specifically, there exist three distinct frequency-dependent motion regimes.  
At very small frequencies, the magnetic moment vector tends to follow the magnetic field, resulting in a \emph{tumbling} motion of the swimmer, in which the body rotates in place but experiences very little net propulsion.  
At slightly larger frequencies, the swimmer begins to exhibit propulsion by precessing about an axis. 
Finally, there exists a second critical frequency beyond which the magnetic moment vector of the swimmer is unable to follow the rotating magnetic field.  
This critical frequency is commonly known as the \emph{step-out} frequency. 
In this regime, swimmer propulsion falls off rapidly and the swimmer dynamics become difficult to predict or control. 
The dependence on the swimmer velocity on the driving frequency is illustrated in Fig. \ref{fig:VelFreq}. 
These distinct regimes have been shown previously for different body geometries, both experimentally and theoretically, in many works, such as \cite{GhoshPRE,Leshansky_prf_2017}.  

In this paper, we will show that while the swimmer does not exhibit meaningful propulsion in the tumbling regime in the tumbling regime, the rotation of the body as the magnetic moment vector follows the magnetic field generates a rotating fluid velocity field that may be useful for stirring the fluid or manipulating particles.  
That is, by choosing a driving frequency in the propulsion regime, the swimmer may be relocated to a different point in the fluid.  
Then, the axis of rotation may be chosen, with a driving frequency below the first critical frequency, so that the swimmer manipulates the fluid in a desired way, without much motion of the artificial swimmer.
It should be noted that the values of the critical driving frequencies separating the different motion regimes depends on the body geometry, magnetization, and the strength of the applied field \cite{Leshansky_prf_2017}.  For the parameter values as given above, the values of these two critical frequencies were numerically found to be approximately $\omega_1 = 1.55$ Hz and $\omega_2 = 2.38$ Hz.

\section{Particle Manipulation}
\subsection{Particle dynamics in the propulsion regime} \label{ss:above}

\begin{figure*}
\centering
\includegraphics[width = 0.32\linewidth]{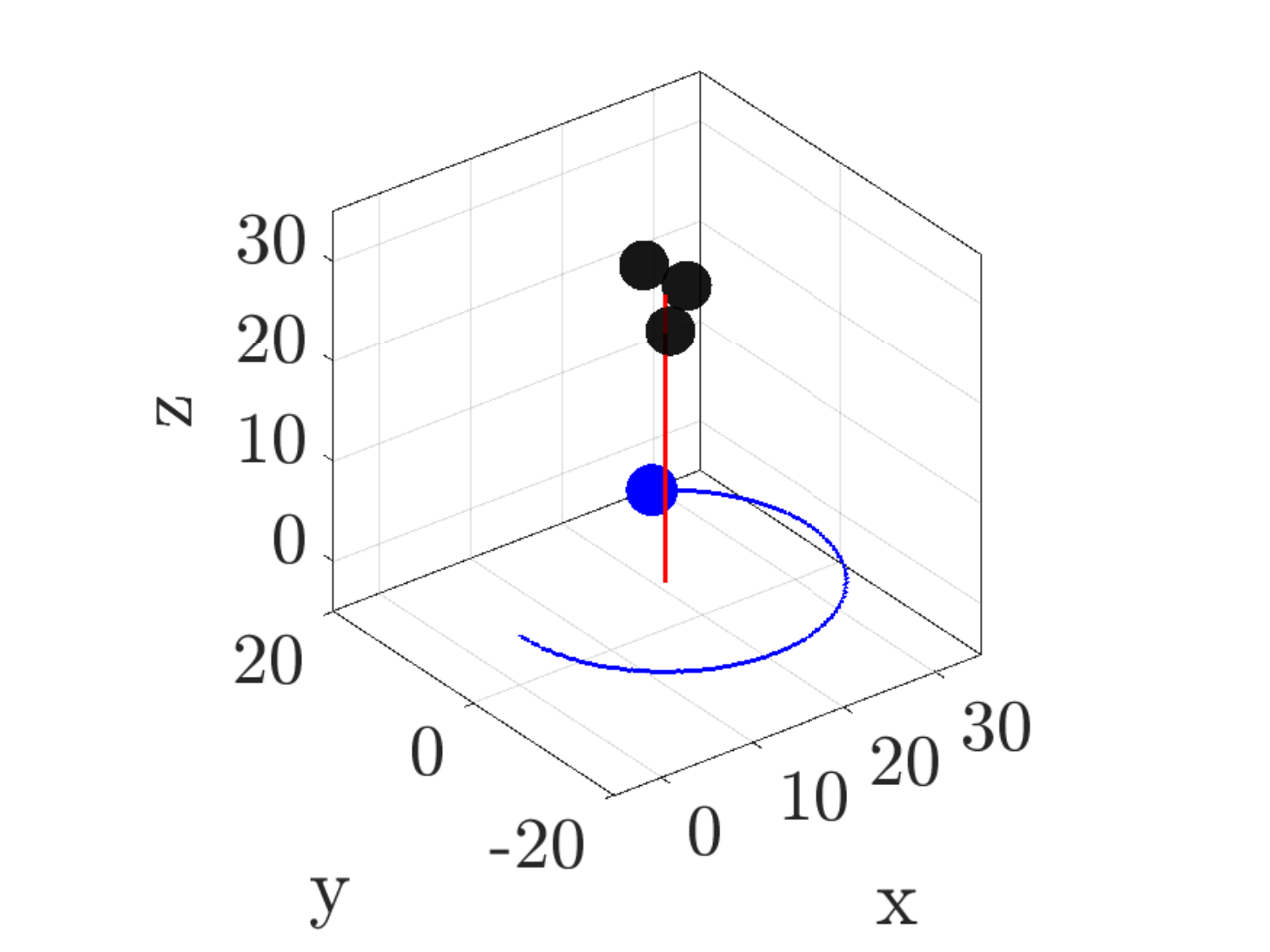}
\includegraphics[width = 0.32\linewidth]{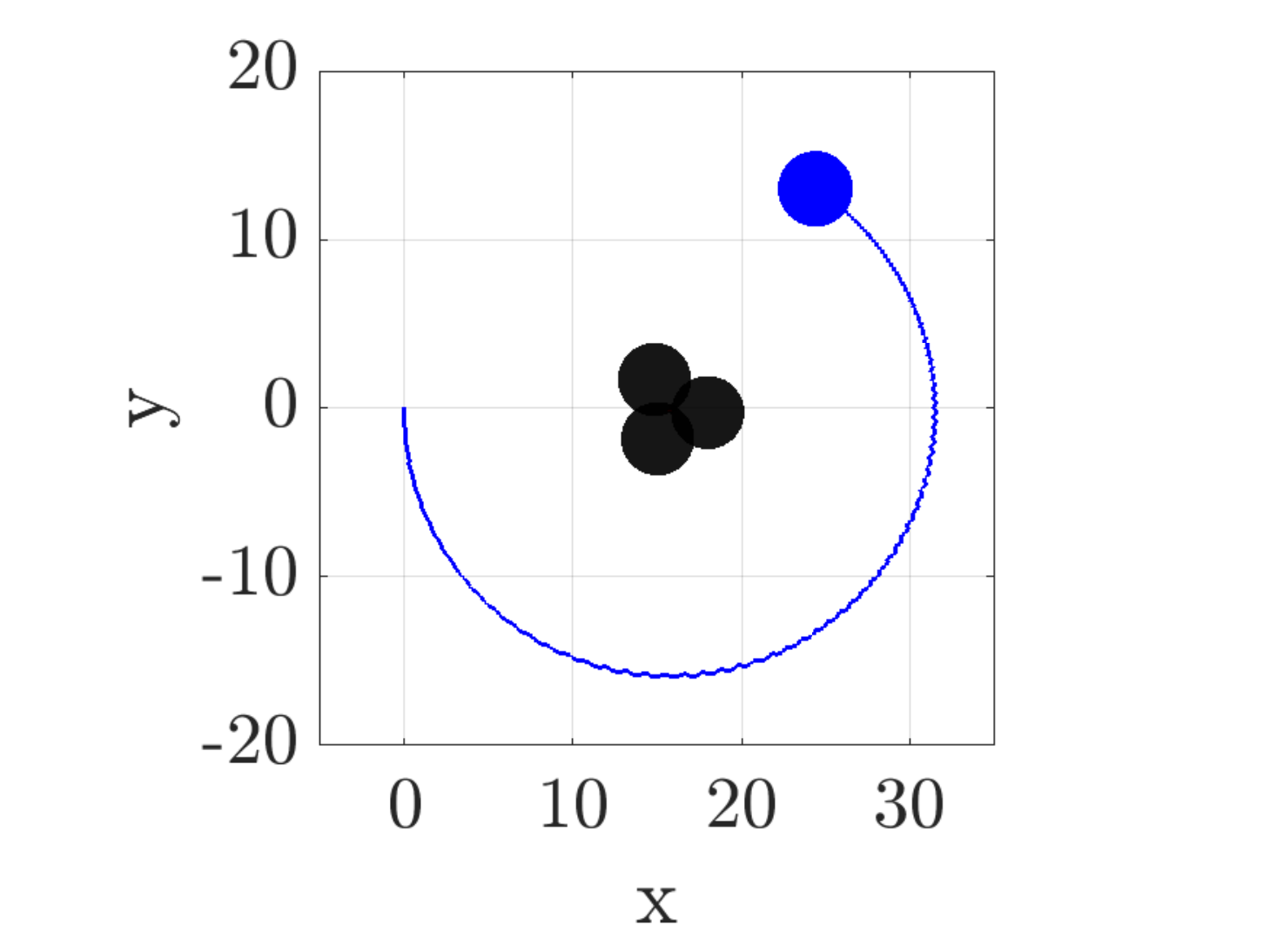}
\includegraphics[width = 0.32\linewidth]{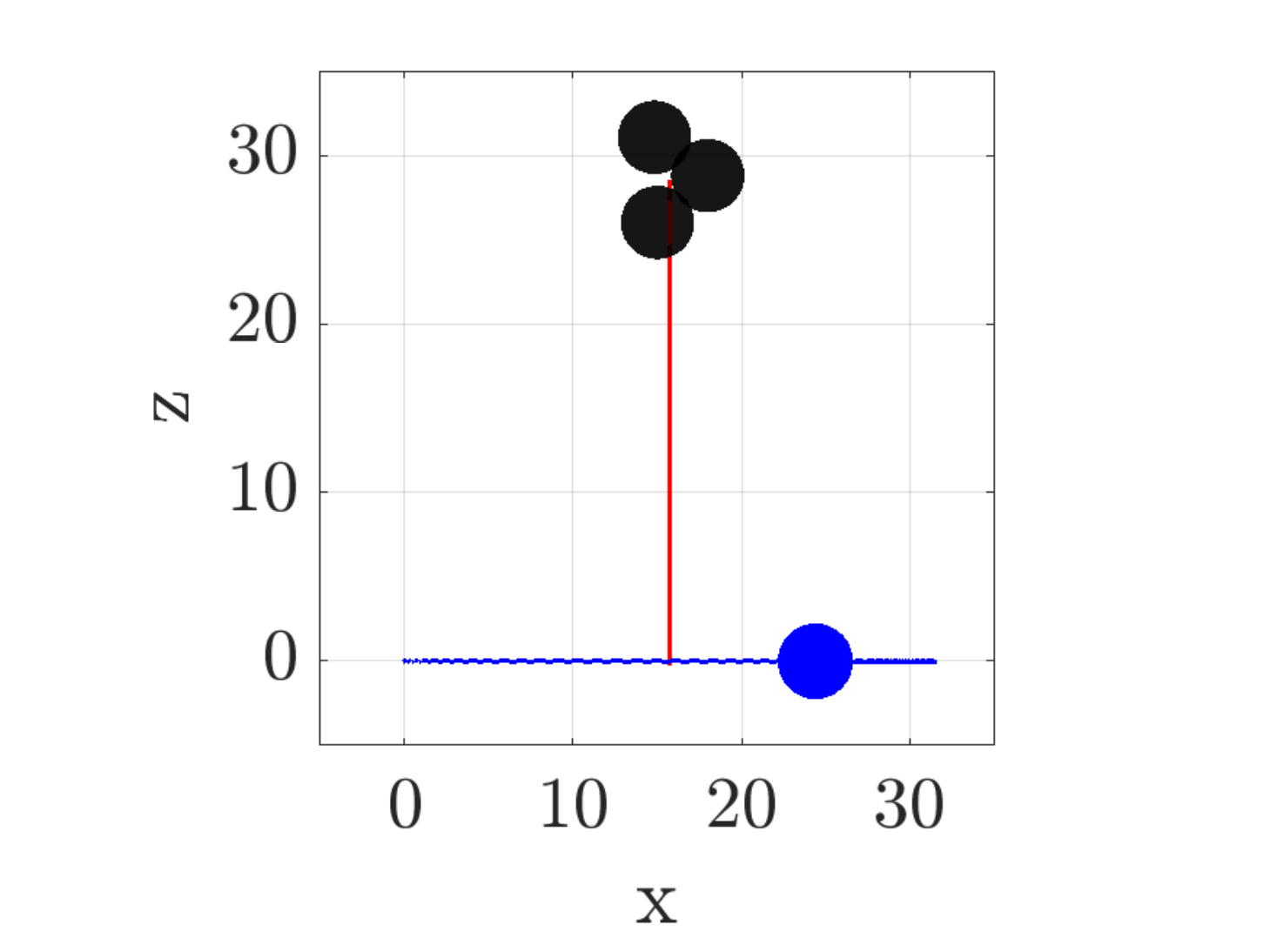}\\
(a)\hspace{6cm} (b)\hspace{6cm} (c)\\
\vspace{1em}
\caption{Trajectories of a magnetically driven three-sphere artificial microswimmer and a passive spherical particle over a timespan of 50 seconds when driven at an input angle $\gamma = 0$ and frequency $\omega = 2.0$ Hz, which falls in the propulsion regime.  Distances shown are given in $\mu$m.}
\label{fig:TrajectoriesAbove}
\end{figure*}

With this understanding of the swimmer dynamics, we now examine the system with a passive spherical particle initially located at the origin, with the artificial swimmer located at an initial position 
\[
\mathbf{X}_s = \begin{pmatrix} 8a\quad   &   0\quad   &   0 \end{pmatrix}^{\intercal}.
\]
The swimmer is then driven by an oscillating magnetic field of the form of Eq \eqref{eq:MagFieldRotY}, with $\gamma = 0$ and $\omega = 2.0$ Hz over a timespan from 0 to 50 s. The resulting swimmer and particle trajectory of this simulation are shown in Fig.  \ref{fig:TrajectoriesAbove}. 

\begin{figure}[H]
    \centering
    \includegraphics[width=0.49\linewidth]{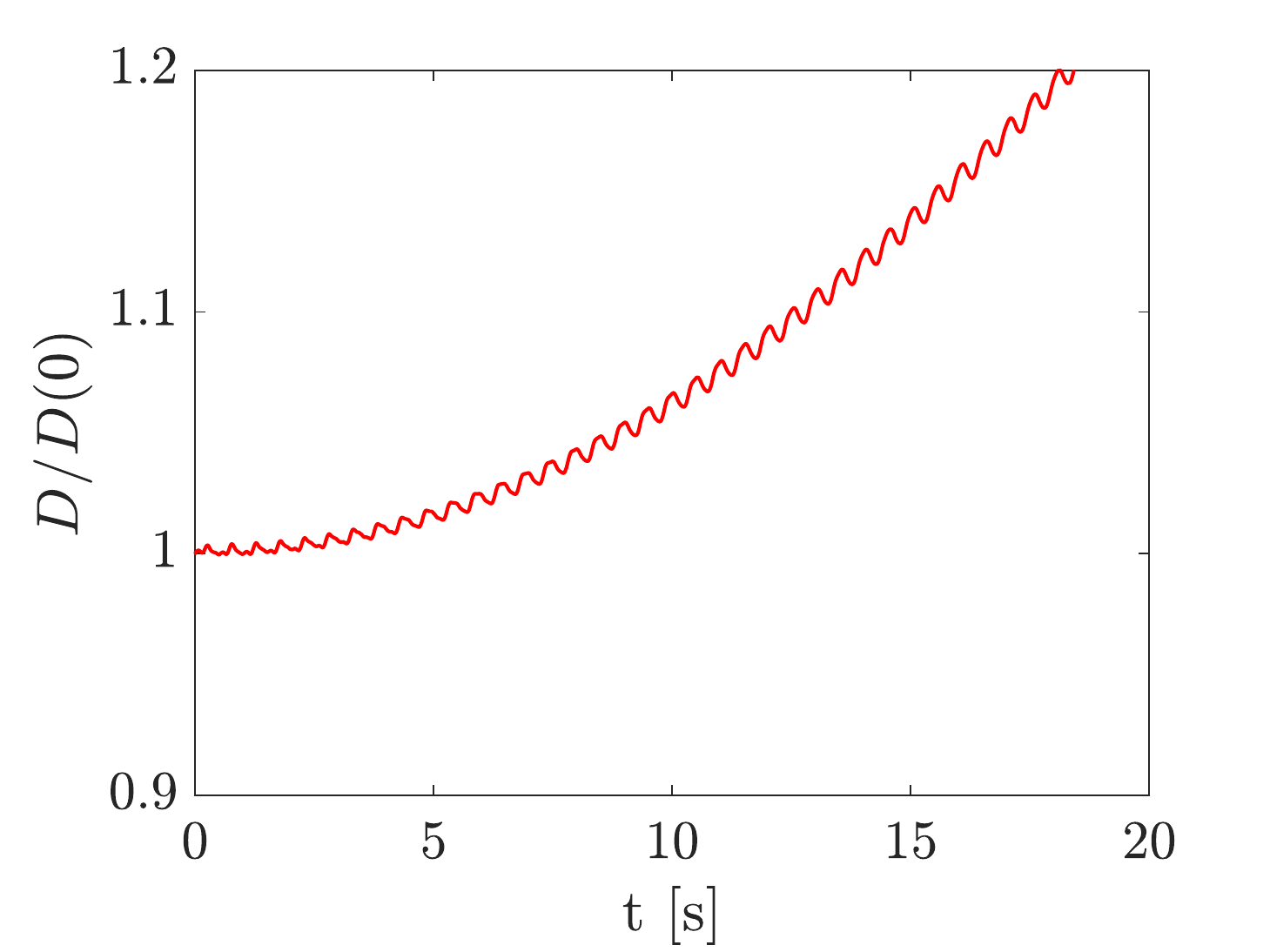}
    \includegraphics[width=0.49\linewidth]{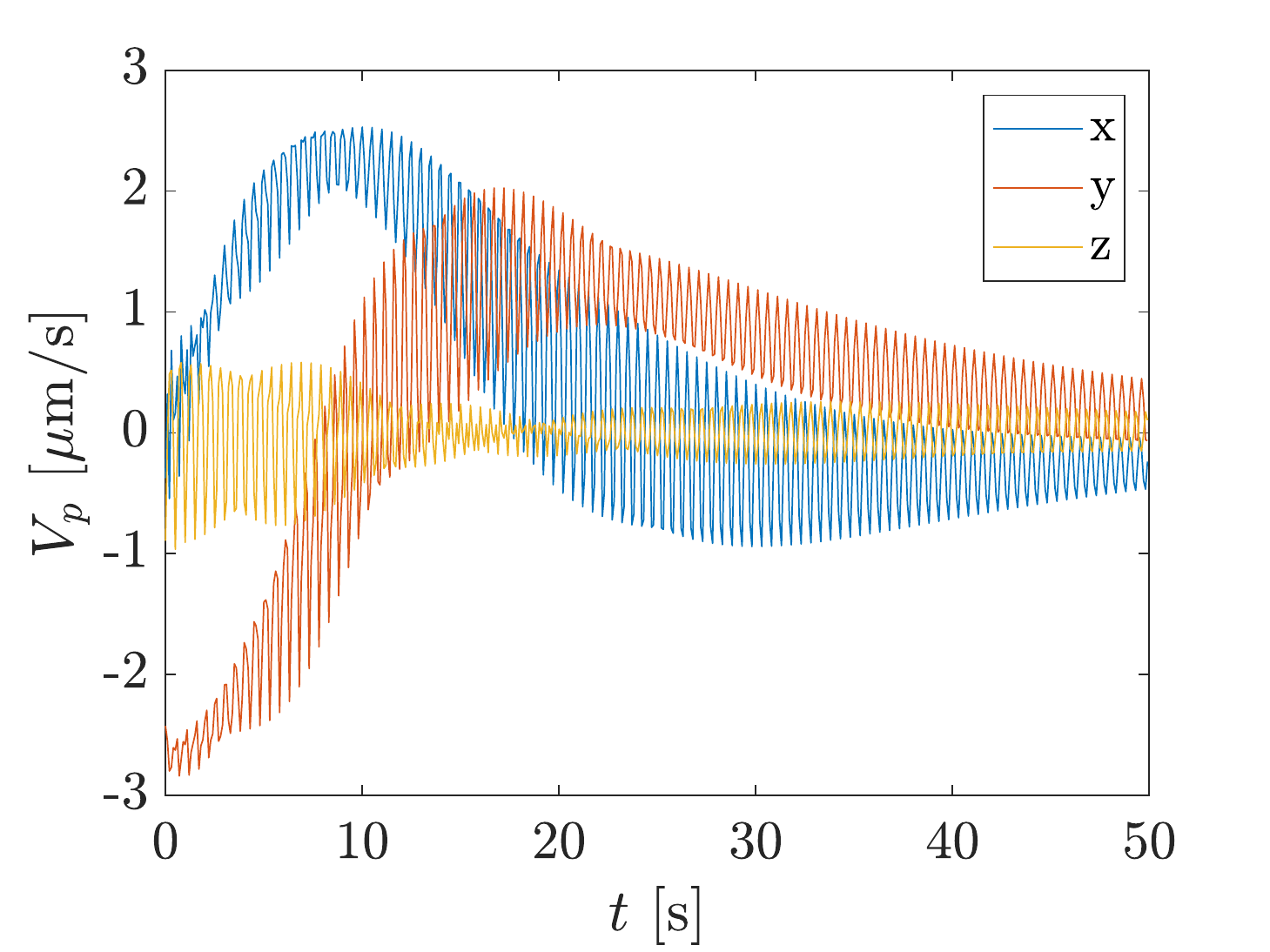}\\
    (a)\hspace{4.5cm}(b)\\
    \vspace{1ex}
    \caption{(a) Distance between the artificial swimmer and the spherical particle as a function of time.  (b) Components of the spherical particle's velocity during the simulation, represented in the spatially-fixed frame of reference. Both plots correspond to the trajectory shown in Fig. \ref{fig:TrajectoriesAbove}.}
    \label{fig:DistVelAbove}
\end{figure}

From these trajectories, it can be seen that the swimmer's rotational motion causes it to translate in the positive $z$ direction.  Furthermore, the rotational velocity field produced by the swimmer causes the passive spherical particle to move in a circular path about the swimmer in the plane orthogonal to the direction of rotation and swimmer motion, with very little motion out of that plane.  For this reason, the swimmer tends to move away from the passive particle, and thus the fluid velocity field produced by the swimmer tends to decay in the vicinity of the particle. This effect is illustrated in Fig. \ref{fig:DistVelAbove}.  
Figure \ref{fig:DistVelAbove} (a) shows that as time progresses, the distance between the swimmer and the sphere grows larger, as the swimmer moves out of the $xy$ plane. Fig. \ref{fig:DistVelAbove}(b) shows the components of particle velocity throughout the simulation. From this, we see that the $z$-component of the particle velocity oscillates about zero for the entirety of the simulation, while the $x$ and $y$ components tend towards zero as the swimmer moves away from the particle. For this reason, it becomes difficult to generate significant particle motion with a swimmer operating in this regime, as the swimmer will typically tend to move away from the particle. 

\subsection{Particle dynamics in the tumbling regime} \label{ss:below}
Seeing that the swimmer tends to move away from the passive particle when operating at a frequency in the propulsive regime motivates us to examine the motion of a particle as the swimmer is driven by a magnetic field rotating at a frequency below the first critical frequency.  As described previously, in this regime, the magnetic moment vector of the swimmer still tends to follow the rotating magnetic field, resulting in a rotating motion of the swimmer body.  However, this motion is not enough to provide the body with a net translation, so the swimmer tends to rotate, or tumble, in place without much net motion. 

\begin{figure*}[ht!]
\centering
\includegraphics[width = 0.32\linewidth]{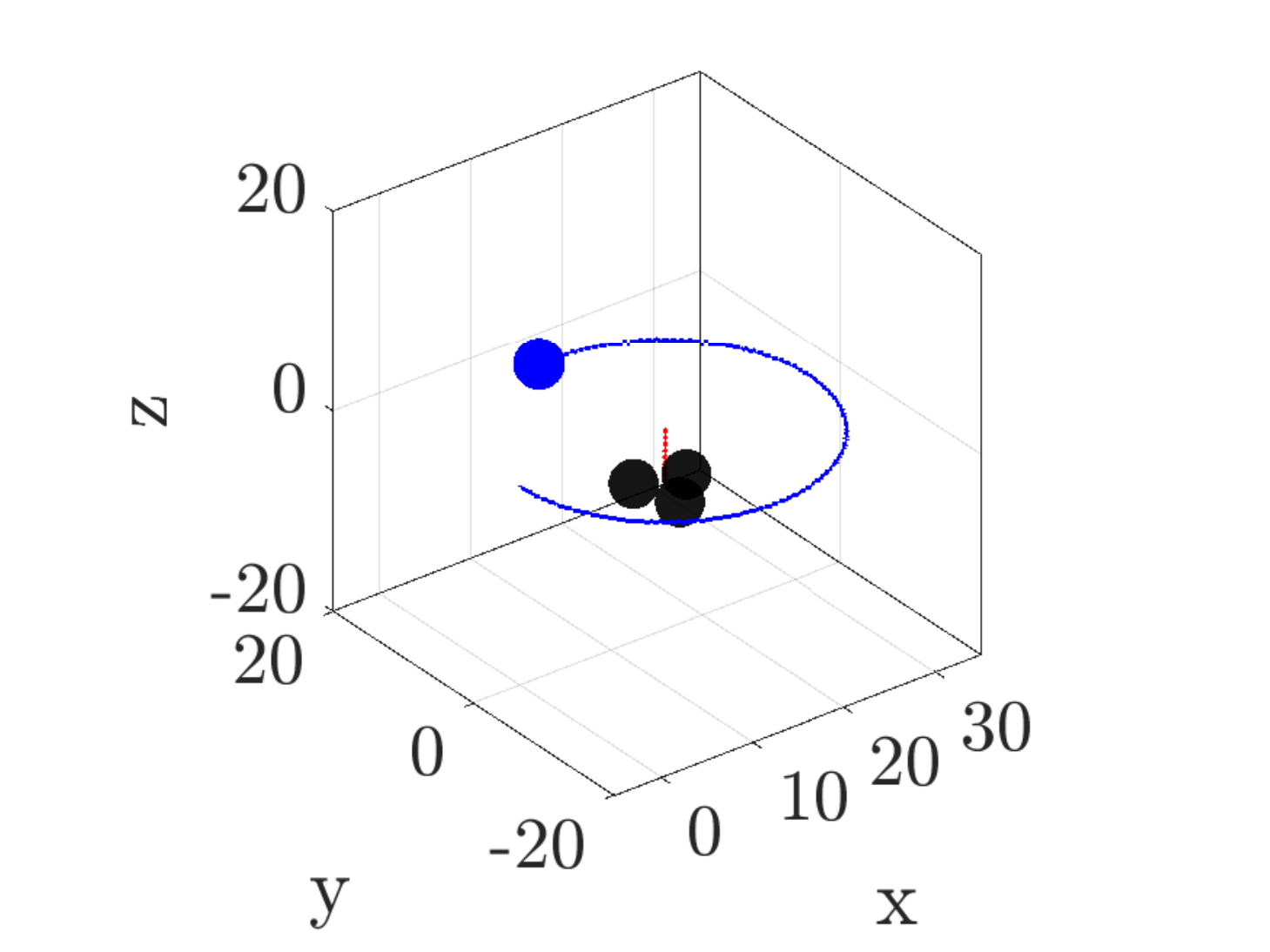}
\includegraphics[width = 0.32\linewidth]{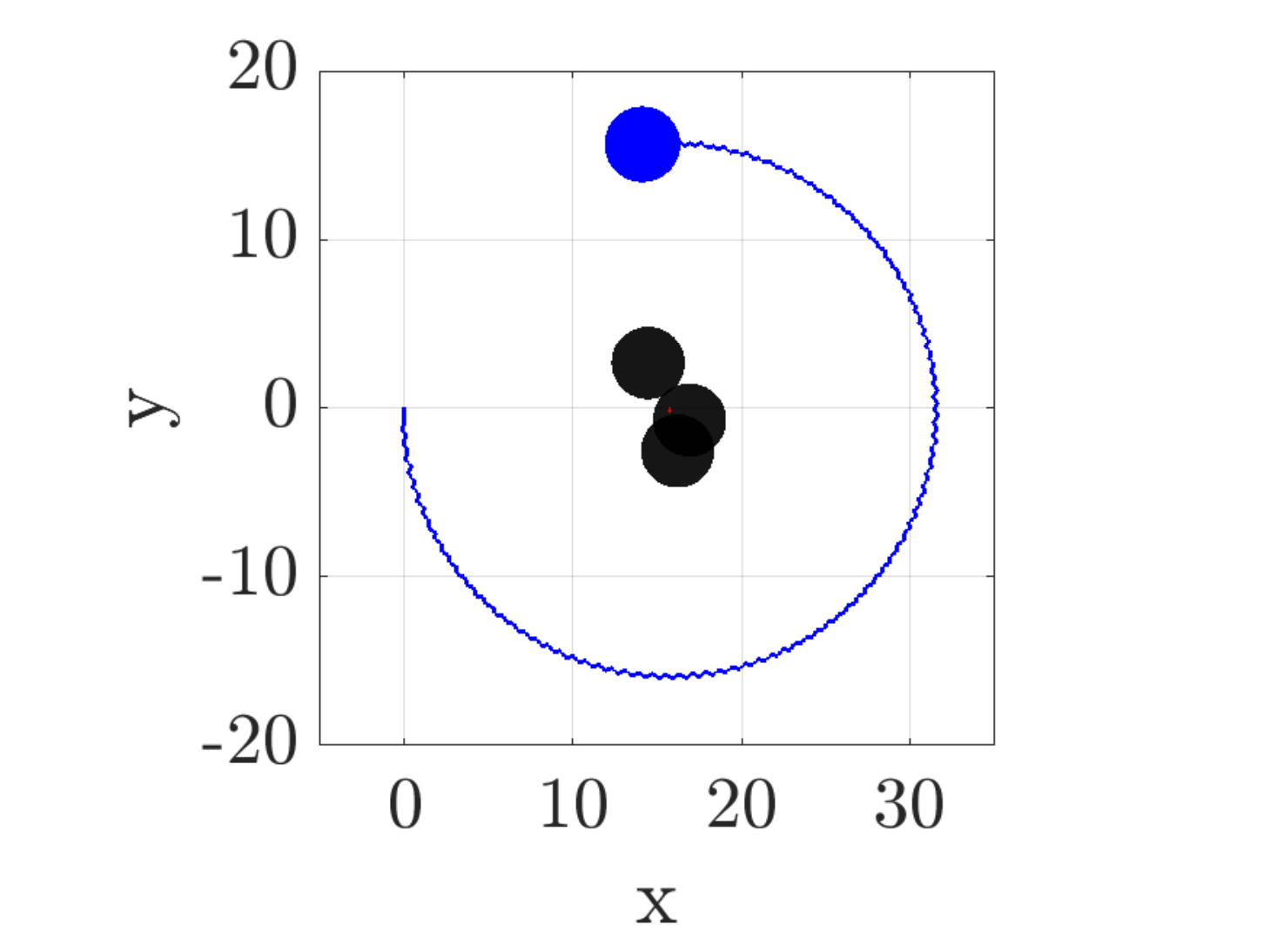}
\includegraphics[width = 0.32\linewidth]{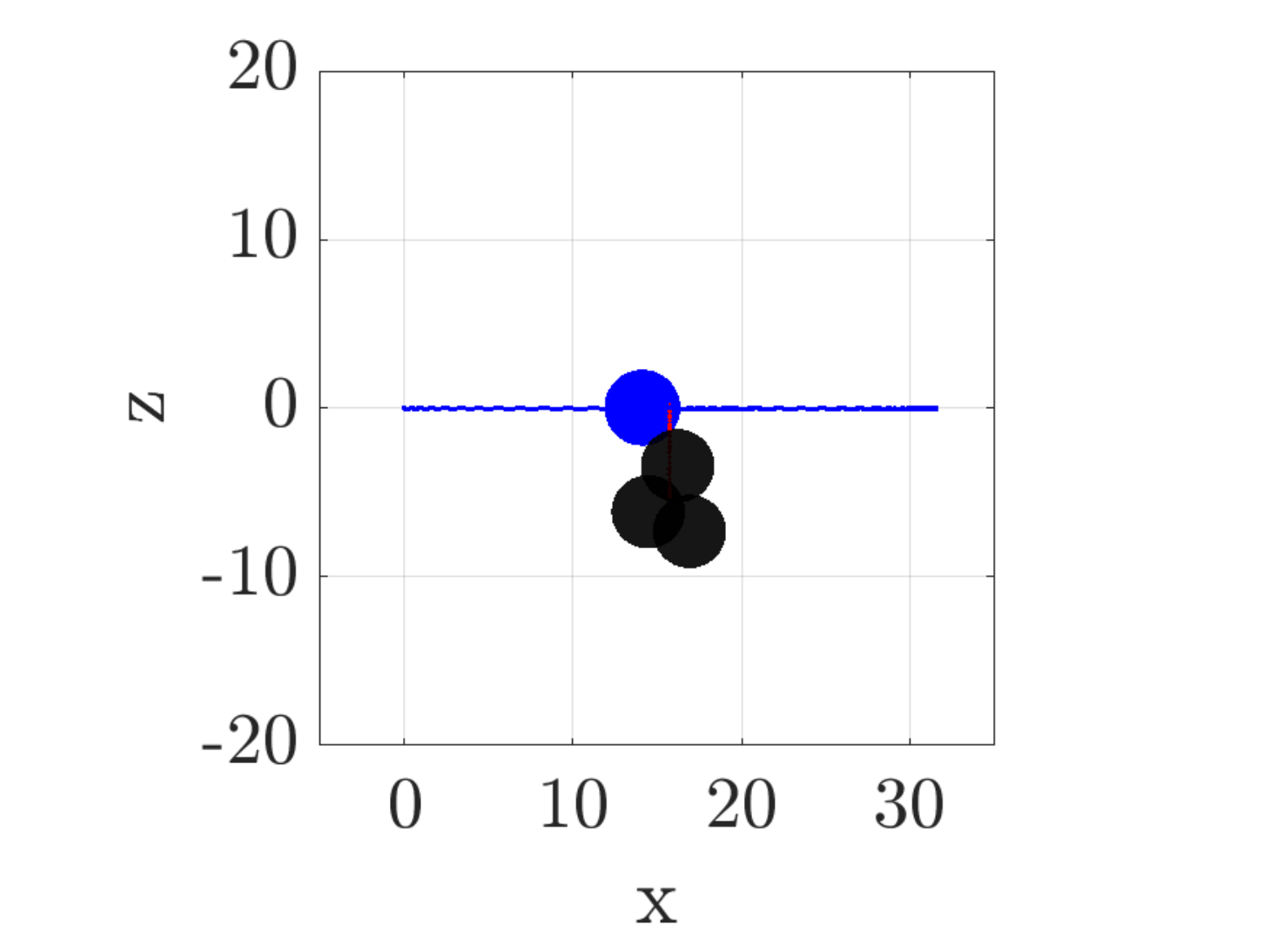}\\
(a)\hspace{6cm} (b)\hspace{6cm} (c)\\
\vspace{1em}
\caption{Trajectories of a magnetically driven three-sphere artificial microswimmer and a passive spherical particle over a timespan of 50 seconds when driven at an input angle $\gamma = 0$ and frequency $\omega = 1.0$ Hz, which falls in the tumbling regime.  Distances shown are given in $\mu$m.}
\label{fig:TrajectoriesBelow}
\end{figure*}

\begin{figure}[H]
    \centering
    \includegraphics[width=0.49\linewidth]{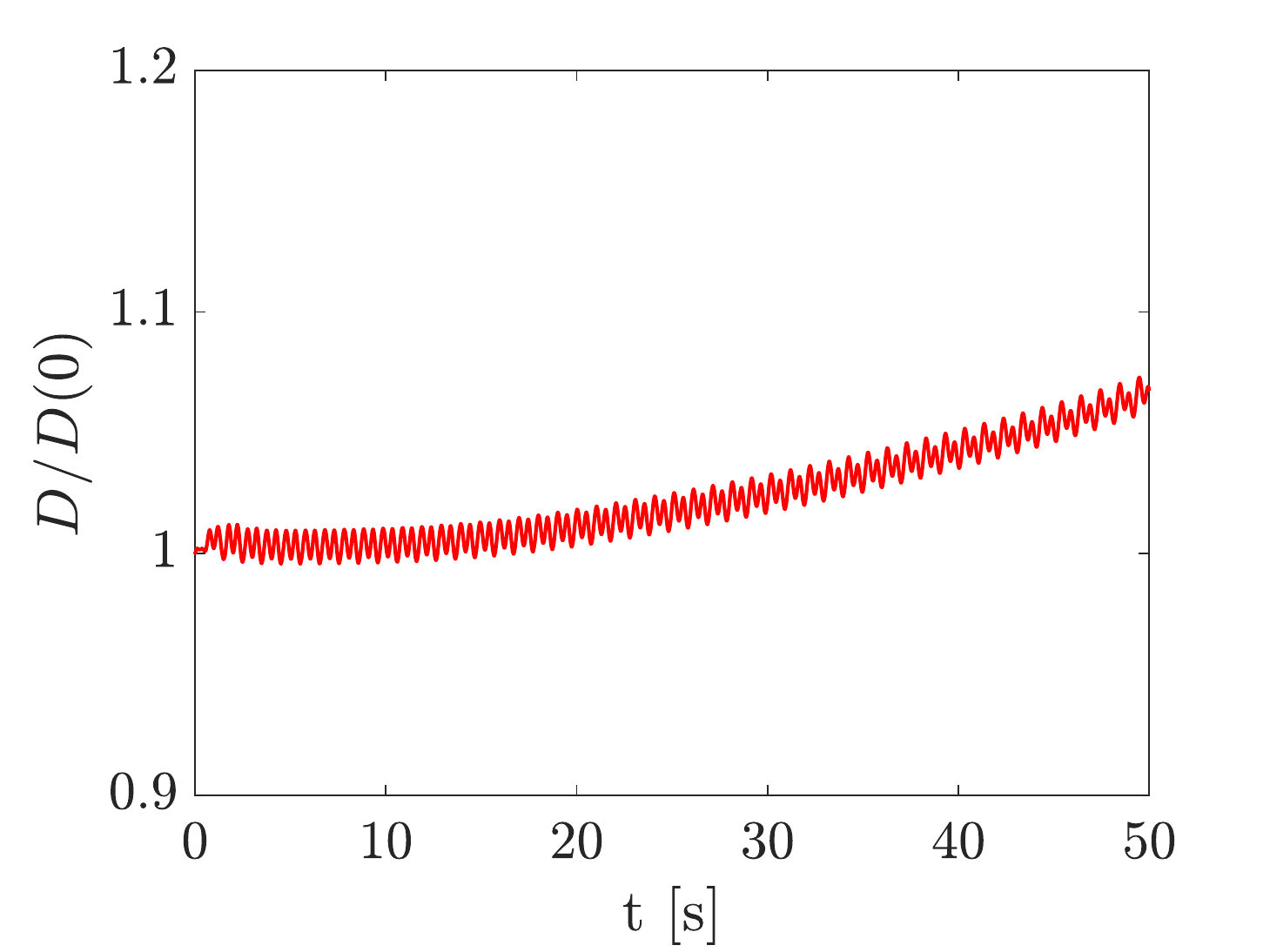}
    \includegraphics[width=0.49\linewidth]{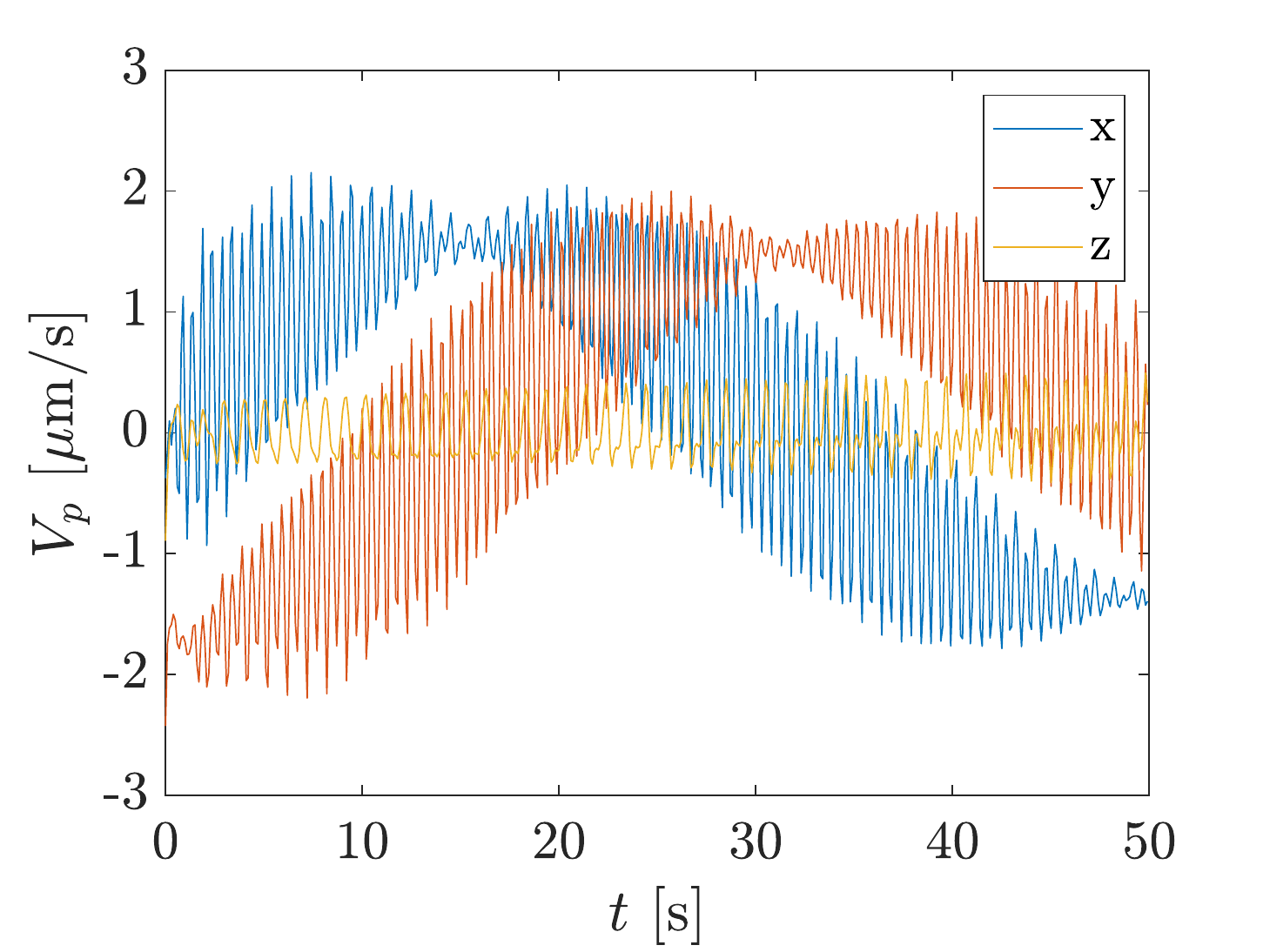}\\
    (a)\hspace{4.5cm}(b)\\
    \caption{(a) Distance between the artificial swimmer and the spherical particle as a function of time.  (b) Components of the spherical particle's velocity during the simulation, represented in the spatially-fixed frame of reference. Both plots correspond to the trajectory shown in Fig. \ref{fig:TrajectoriesBelow}}
    \label{fig:DistVelBelow}
\end{figure}

To contrast the tumbling/spinning motion of the microswimmer with the translational motion, a simulation is done with the same initial condition as in the simulation described in \S \ref{ss:above}, still with a magnetic field angle of $\gamma=0$, but with a driving frequency of $\omega = 1.0$ Hz. This falls in the tumbling regime, as can be seen in Fig. \ref{fig:VelFreq}.  The resulting trajectories over a 50 second simulation with this input are shown in Fig. \ref{fig:TrajectoriesBelow}.  The corresponding distance between the particle and the swimmer are plotted and shown in Fig. \ref{fig:DistVelBelow}.

From the trajectory, it can be seen that the swimmer rotates nearly in place, with only a small amount of translation out of the $xz$ plane.  
The rotational motion of the artificial swimmer's body generates a rotational velocity field in the fluid that is similar to that generated by a rotlet. This rotational velocity of the fluid causes the passive spherical particle to translate in the $xy$-plane.  
Comparing the trajectory in Fig. \ref{fig:TrajectoriesAbove} to the trajectory of Fig. \ref{fig:TrajectoriesBelow}, it can be seen that the sphere travels farther along its circular arc for the case of the swimmer in the tumbling regime. 
The reason behind this is as follows. 
When the swimmer and the particle lie in the same plane (the $xy$-plane here), the sphere's velocity is higher when the swimmer moves in the propulsive regime than when it moves in the tumbling regime.  
When the swimmer moves in the propulsive regime, it tends to translate out of this plane and away from the particle, whereas in the tumbling regime, the swimmer stays near to this plane throughout the simulation. 
For this reason, the velocity of the sphere does not decay nearly as rapidly in the tumbling case, as the distance between the particle and the sphere grows much more slowly.  
Both of these findings can be observed in Fig \ref{fig:DistVelBelow}.

\begin{figure*}[ht]
\includegraphics[width = 0.32\linewidth]{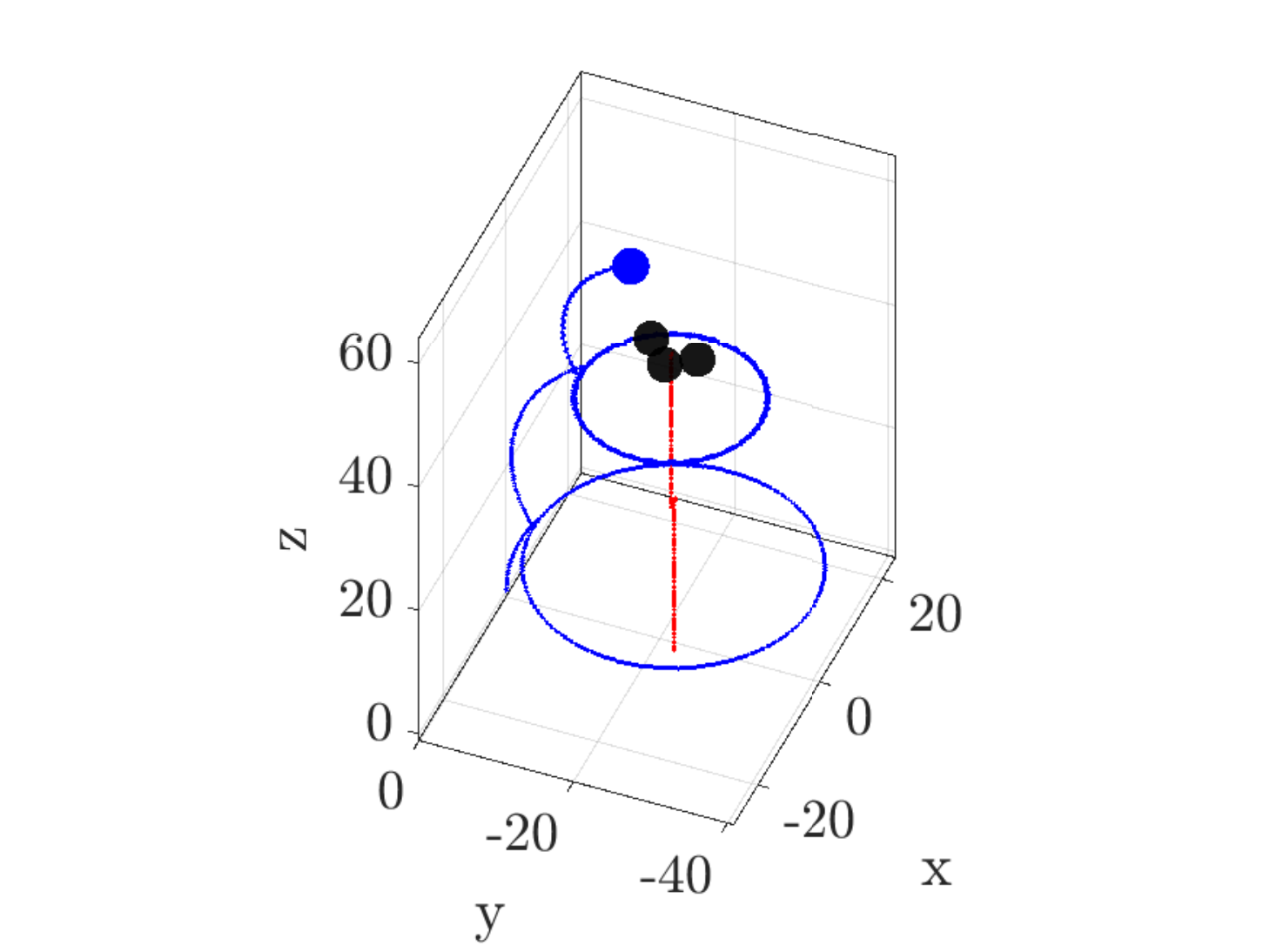}
\includegraphics[width = 0.32\linewidth]{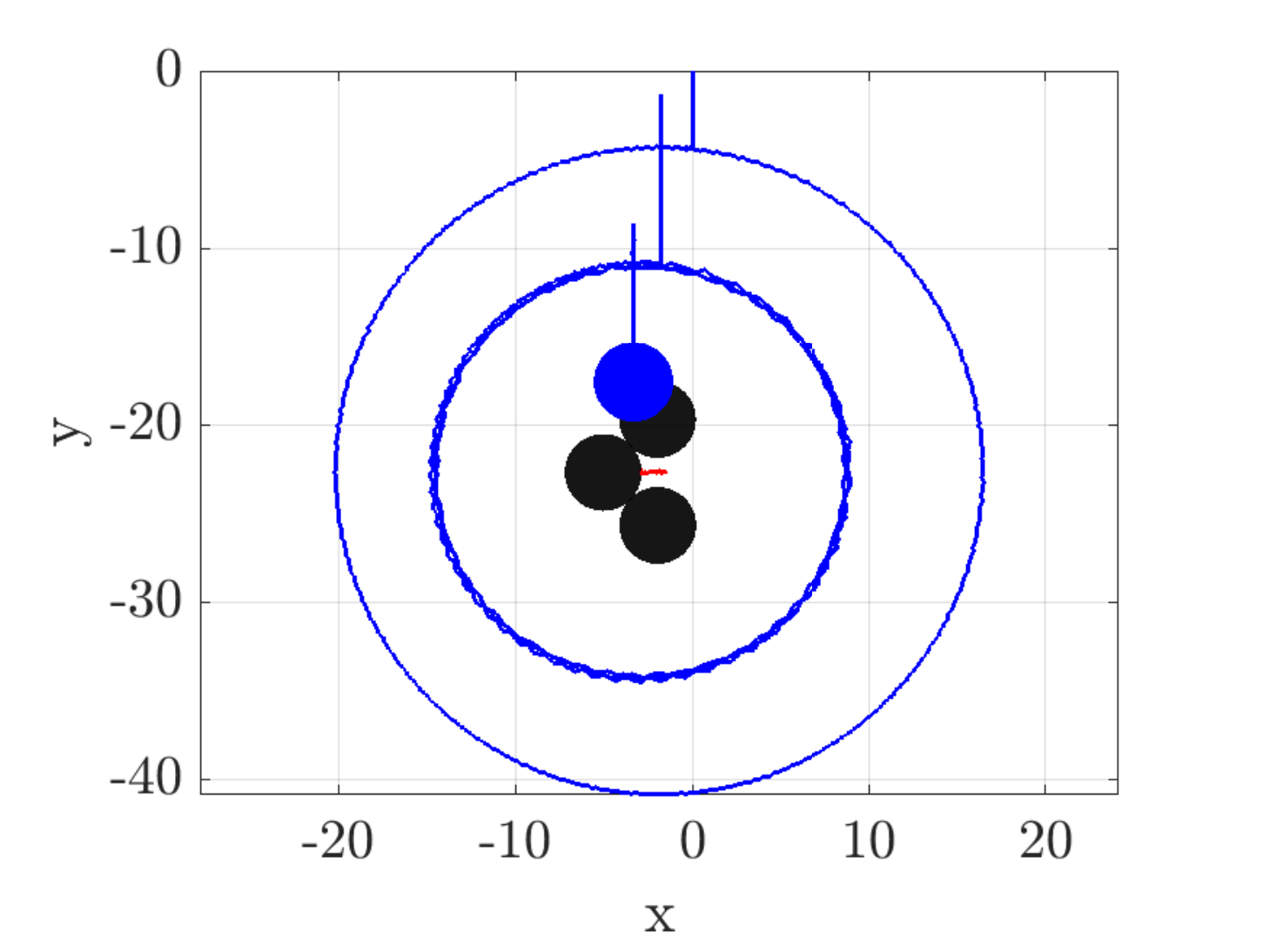}
\includegraphics[width = 0.32\linewidth]{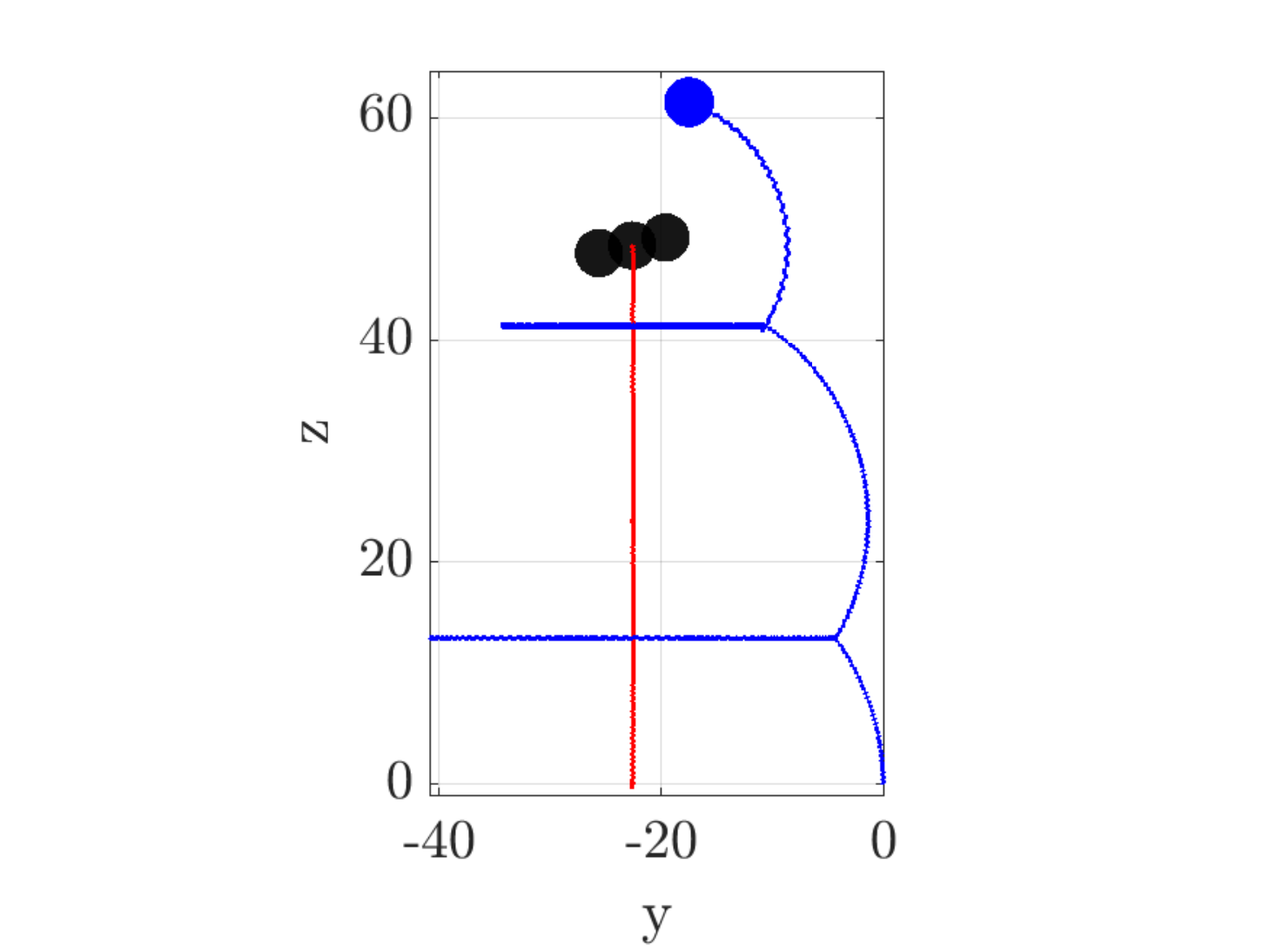}
\caption{Trajectory of the magnetically driven swimmer and passive spherical particle in response to a control input that switches between piecewise constants, as described in \S\ref{ss:ControlPiecewise}.  Distances shown are given in $\mu$m. }
\label{fig:TrajectoriesSwitched}
\end{figure*}
\subsection{Control by piecewise steady inputs} \label{ss:ControlPiecewise}
In the previous two simulations shown, neither of the applied inputs resulted in consistent, significant propulsion for both the passive particle and the artificial swimmer. When the swimmer is driven in the tumbling regime, the particle exhibits significant motion, moving in a circular arc about the swimmer.  However, not much net translation is achieved, as the particle tends to return to its original position.  In the propulsive regime, the swimmer exhibits significant propulsion, translating in the direction of magnetic field rotation, but it tends to move away from the particle, resulting in small particle displacement.  With this in mind, we seek to consider a sequence of driving frequencies $\omega$ and directions $\gamma$ to steer both the swimmer and particle in a desired direction. 

To accomplish this, the task is broken into phases in which the swimmer is to be moved and phases in which the passive particle is to be moved.  
To move the swimmer, the magnetic field's axis of rotation is oriented along the desired direction and a field rotation frequency should be chosen below the step-out frequency, but above the tumbling-propulsion transition frequency. 
To move the passive particle, the magnetic field is oriented in a direction perpendicular to the desired direction of motion, and a frequency is chosen to place the swimmer in the tumbling regime. 
To illustrate this, we simulate the system with a passive particle initially located at the origin and the artificial swimmer initially located at the position 
\[
\mathbf{X}_s = \begin{pmatrix} 0\quad   &   -10a\quad   &   0 \end{pmatrix}^{\intercal}.
\]
With this, we choose a sequence of inputs $\gamma$ and $\omega$ in a piecewise constant manner with the goal of steering the swimmer in the positive $z$ direction of the spatially fixed reference frame.  
More specifically, for one interval of time, we choose an input of $\omega = 1$ Hz so that the swimmer will undergo negligible translational motion and $\gamma = \pi/2$ rad, so that the magnetic field axis of rotation is aligned with the positive $x$-direction of the spatially-fixed frame.  
This rotation causes the particle to move in a circular arc about the swimmer in the $yz$-plane.
This motion is then halted before the sphere reaches the top of this circle and begins to undergo motion in the negative $z$-direction. 
Once this motion has completed, a new control input is selected and held constant over a time interval, this time chosen with the objective of moving the swimmer in the positive $z$-direction.  
For this, the inputs $\omega = 2$ Hz and $\gamma = 0$ rad/s are chosen, so that the artificial swimmer moves in the propulsive regime along an axis oriented in the positive $z$-direction. 
This motion is carried out until the artificial swimmer moves past the spherical particle in the positive $z$ direction, but not so long that the particle lies outside of the range of influence of the artificial swimmer's velocity field.  
At the completion of these two time intervals, both the swimmer and the particle will have achieved a net translation in the positive $z$ direction.  This cycle can thus be applied iteratively to achieve a larger net displacement.  
A trajectory resulting from such an input is generated and shown in Fig. \ref{fig:TrajectoriesSwitched}.  
The inputs $\gamma$ and $\omega$ applied here are as described above and as displayed below in Fig. \ref{fig:SwitchedInputs}.  
Here, the lengths of time intervals are chosen somewhat arbitrarily, and differ in order to illustrate differences in the resulting motions. 
 
 \begin{figure}[H]
     \centering
     \includegraphics[width = 0.49\linewidth]{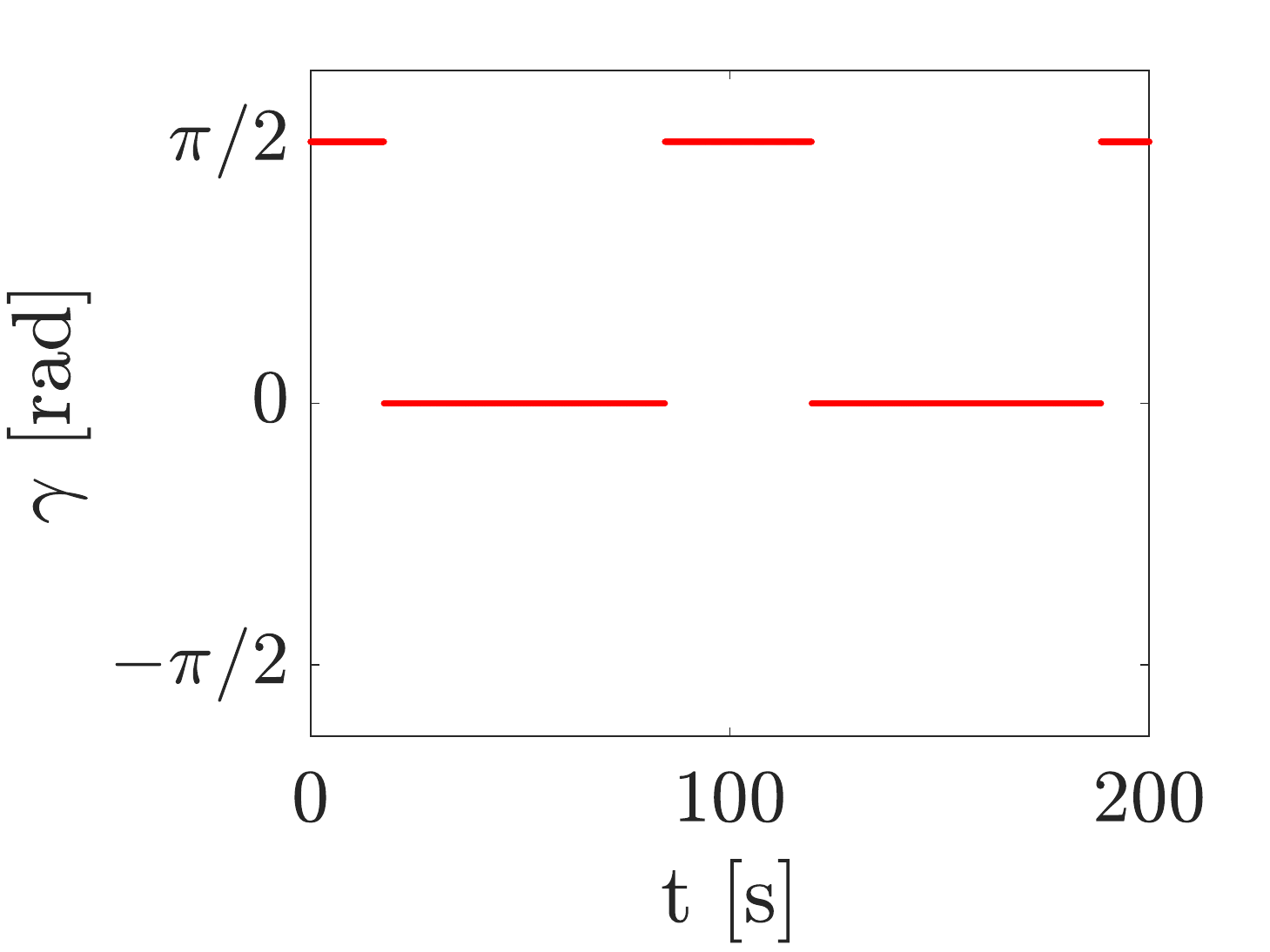}
     \includegraphics[width = 0.49\linewidth]{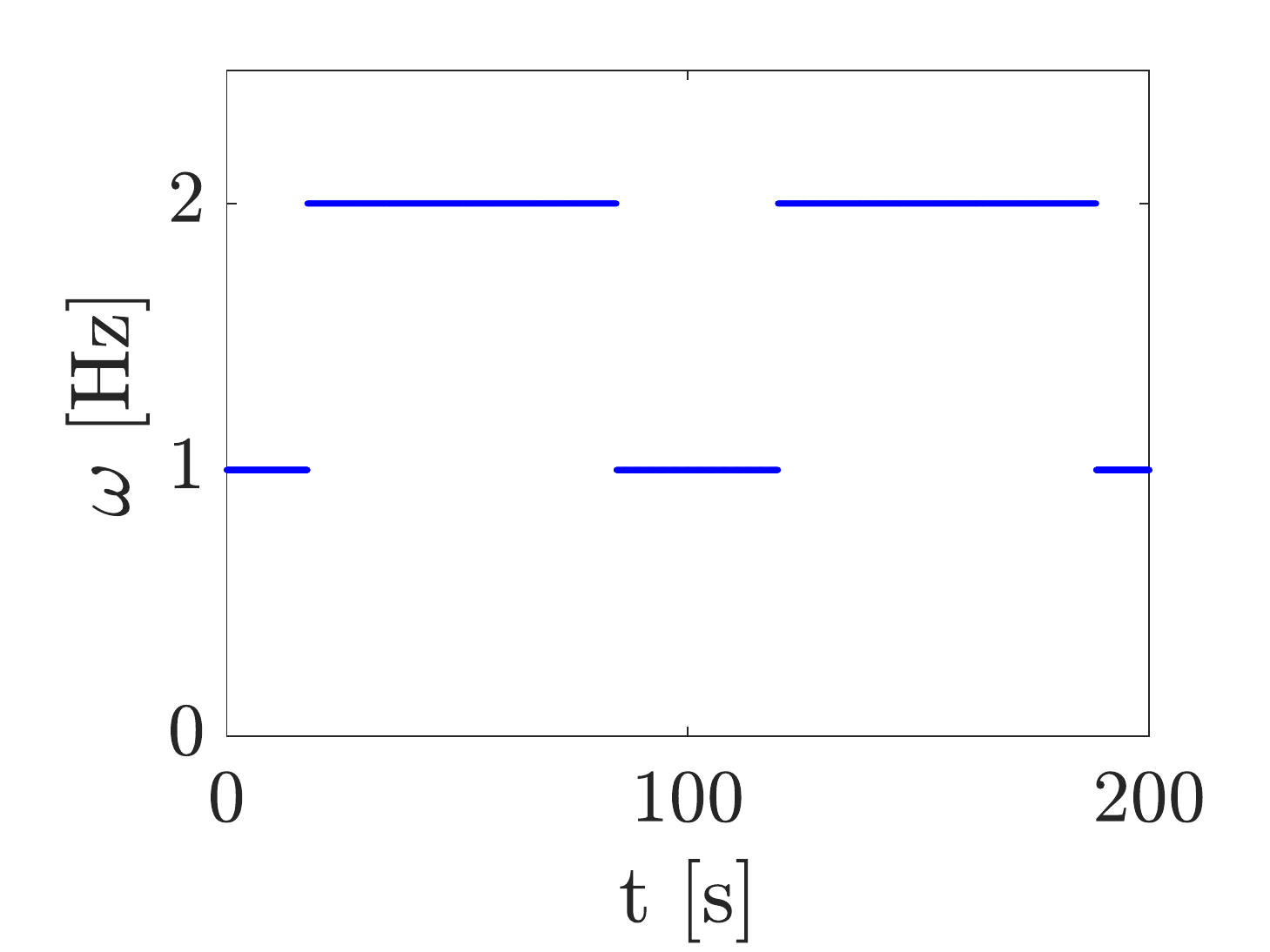}\\
     \caption{Control inputs $\omega$ and $\gamma$ corresponding to the trajectories shown in Fig. \ref{fig:TrajectoriesSwitched}.}
     \label{fig:SwitchedInputs}
 \end{figure}
 
\begin{figure}[!h]
     \centering
     \includegraphics[width = 0.49\linewidth]{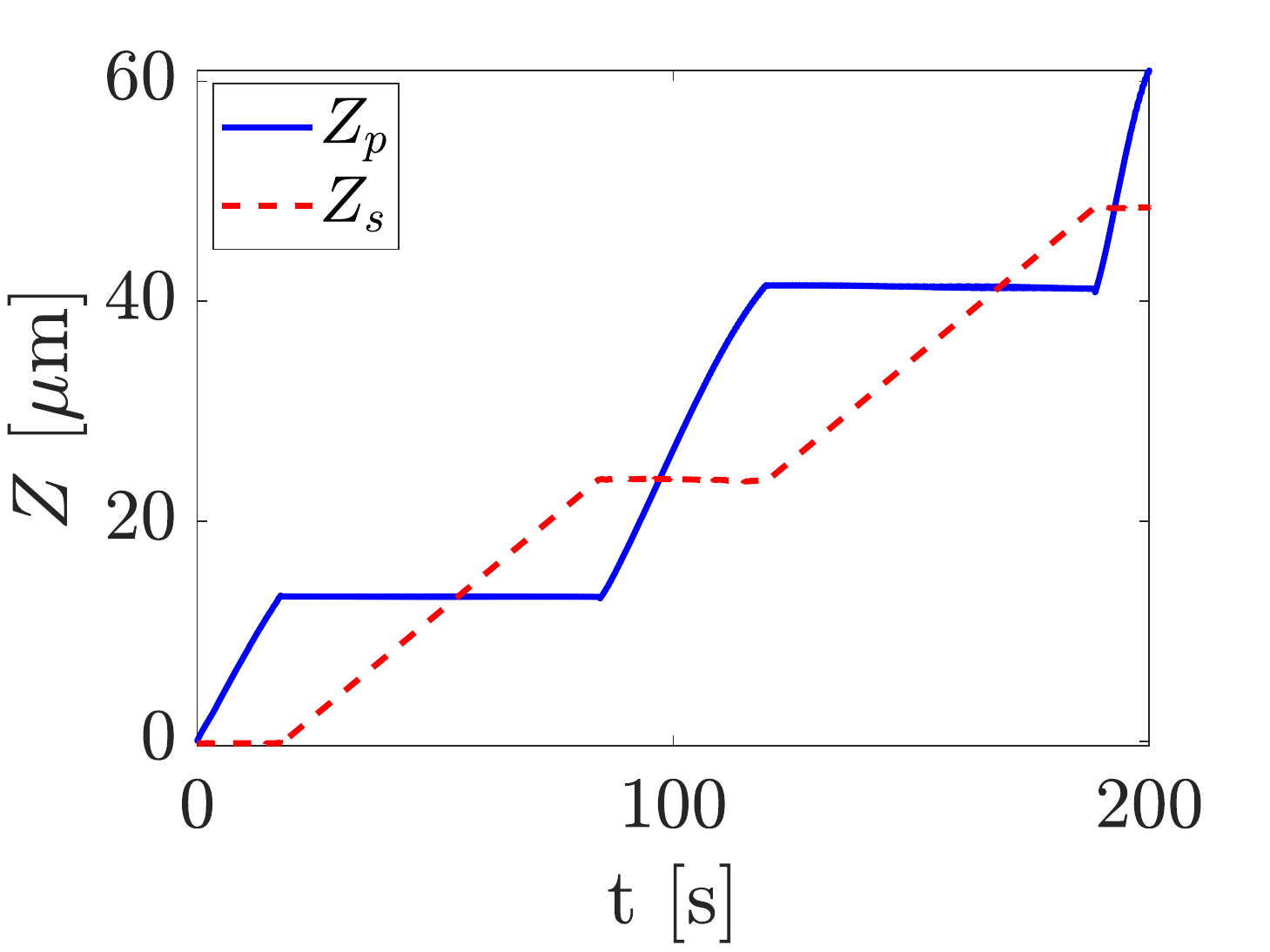}
     \includegraphics[width = 0.49\linewidth]{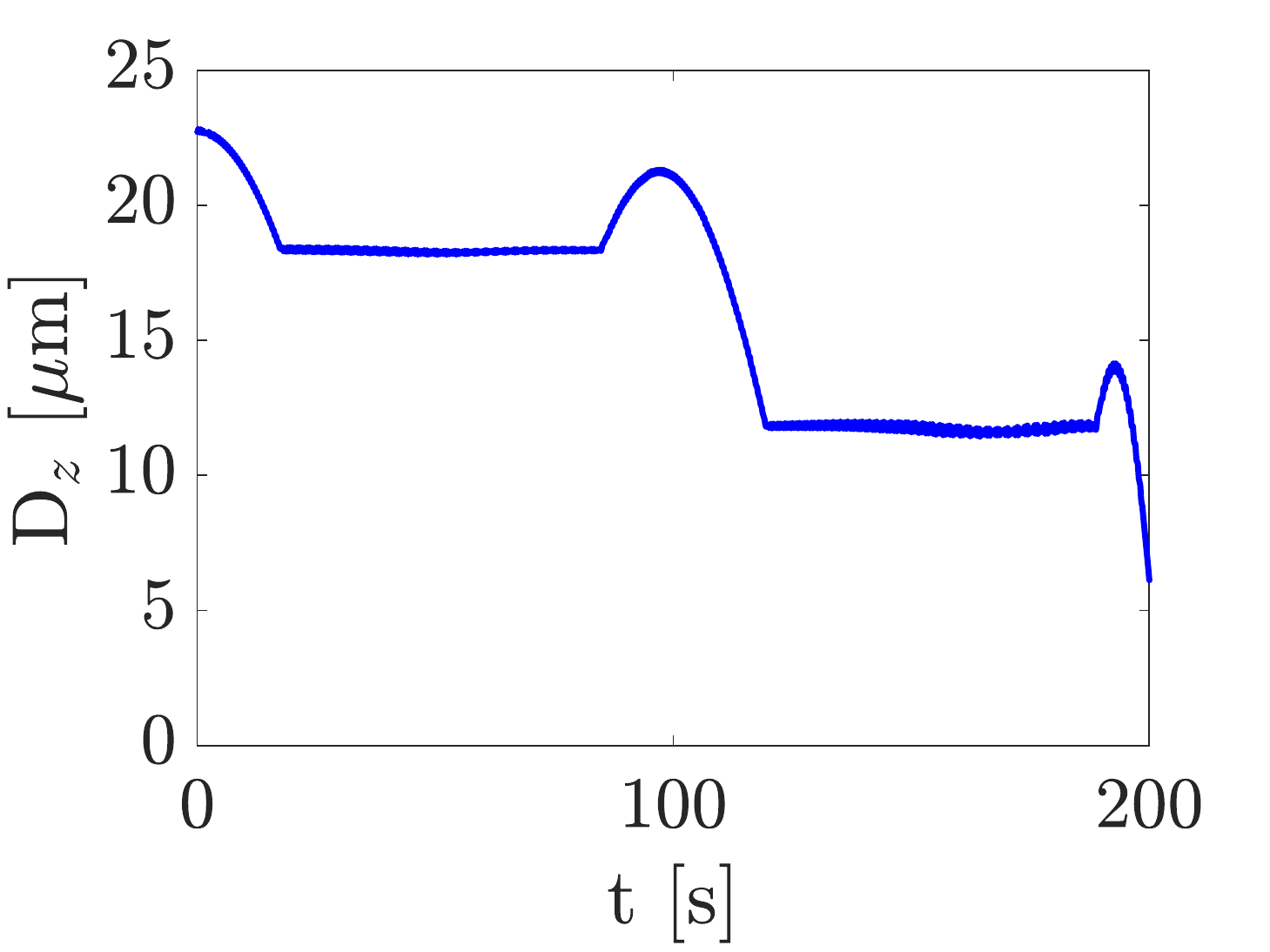}\\
     \caption{(a) Position in the $z$ direction of the swimmer and the particle respectively over time. (b) Distance $D_z$ from the desired direction of swimmer motion. Both plots correspond to the simulation described in \S \ref{ss:ControlPiecewise}.}
     \label{fig:ZAndDz}
 \end{figure}

 This simulation shows that translation of both the particle and the artificial swimmer in the positive $z$-direction can be achieved using the sequence of inputs as described above.  This is emphasized in Fig. \ref{fig:ZAndDz} (a) below, which shows the $z$ position of both bodies as a function of time throughout the simulation.  This demonstrates precisely how each of the control inputs leads to a translation of the swimmer or the sphere, respectively.  Furthermore, since the choice of translation in the $z$-direction was arbitrary in this simulation, this indicates that motion of these bodies may be achieved in any direction by using a comparable algorithm.  

 \vspace{-5pt}
  While this simulation clearly demonstrates that the particle and the swimmer may be moved together, it also shows that the distance between the particle and the swimmer's axis of rotation may be changed. That is, there exists a sequence of inputs that allow the particle to be driven radially inwards or outwards from the swimmer.  This can be seen by examining the distance of the particle from the axis representing the desired propulsion direction.  Here, we denote this distance by $D_z$ and show it in Fig \ref{fig:ZAndDz} (b) as a function of time.   While the trajectories in Fig. \ref{fig:TrajectoriesAbove} and \ref{fig:TrajectoriesBelow} show that this in-plane distance typically remains constant for a constant input, we have shown that by combining these inputs in a piecewise constant manner, this radial distance can be reduced.
  Further, due to the linearity and time-reversibility of low Reynolds number locomotion, this indicates that the radial distance may also be enlarged by applying the time-reversed version of this input.

\section{Conclusion}

Artificial microswimmers have many potential applications in the biomedical field.  
Magnetically driven microswimmers are of particular interest, as they present the capability of being controlled remotely.  
In this work, we examine how a magnetically actuated artificial microswimmer may be used to contactlessly manipulate a passive spherical particle with a size quantified on a comparable length scale.
We have outlined an algorithm for controlling the motion of the passive particle by adjusting the orientation and frequency of the magnetic field's rotation.  
It has been shown that motion of both the swimmer and the a passive spherical particle may be achieved in an arbitrary direction through a sequence of magnetic field inputs in which the direction and frequency are held constant for fixed intervals of time.  
Our simulations also show that the particle may be driven radially inwards or outwards from the artificial swimmer through certain sequences of inputs.  
These results imply that such a swimmer and particle may be driven to any arbitrary position and configuration using the methods proposed herein. 
Therefore, these results imply controllability of the system and thus may be extended and implemented to achieve more complex path planning goals.  While in the past the controllability of idealized systems of singularities was investigated, for instance in \cite{or_siad_2011, bfst_dscc_2018, bft_ijira_2018}, this paper is one of the first to investigate the controllability of the interaction of a cargo particle via the locomotion of a realistic artificial microswimmer.

\bibliographystyle{unsrt}
\bibliography{magswimmer}

\end{document}